\documentclass[twocolumn]{IEEEtran}
\usepackage{authblk}
\usepackage{silence}
\usepackage[font=scriptsize]{caption}
\usepackage{xcolor}
\usepackage{amssymb}
\usepackage{algpseudocode} 
\usepackage{soul}
\usepackage{array}

\usepackage{amsmath,amssymb,amsfonts}

\usepackage{graphicx}
\usepackage{textcomp}
\usepackage{multirow}
\usepackage{program}
\usepackage{hyperref}

\usepackage{cite}
\usepackage{booktabs}
\usepackage{bigstrut}
\usepackage{floatrow}
\usepackage{rotating}
\usepackage{adjustbox}
\usepackage{enumerate}

\usepackage{pifont}

\def\BibTeX{{\rm B\kern-.05em{\sc i\kern-.025em b}\kern-.08em

    T\kern-.1667em\lower.7ex\hbox{E}\kern-.125emX}}
    
\begin{document}

\title{EDDense-Net: Fully Dense Encoder Decoder Network for Joint Segmentation of Optic Cup and Disc}

\author[1,3]{Mehwish Mehmood}
\author[1,2,*]{Khuram Naveed}
\author[4]{Khursheed Auragzeb}
\author[1]{Haroon Ahmed Khan}
\author[4]{Musaed Alhussein}
\author[1]{Syed S. Naqvi}

\affil[1]{Department of Electrical and Computer Engineering, COMSATS University Islamabad, Islamabad, Pakistan}
\affil[2]{Department of Electrical and Computer Engineering, Aarhus University, Aarhus, Denmark}
\affil[3]{The School of Electronics, Electrical Engineering and Computer Science, Queen’s University of Belfast, Belfast, United Kingdom}
\affil[4]{Department of Computer Engineering, College of Computer and Information Sciences, King Saud University, Riyadh, Saudi Arabia}
\affil[*]{Corresponding author: Khuram Naveed, Email: knaveed@ece.au.dk}
\maketitle


\begin{abstract}
Glaucoma is an eye disease that causes damage to the optic nerve, which can lead to visual loss and permanent blindness.
Early glaucoma detection is therefore critical in order to avoid permanent blindness.
The estimation of the cup-to-disc ratio (CDR) during an examination of the optical disc (OD) is used for the diagnosis of glaucoma.
In this paper, we present the EDDense-Net segmentation network for the joint segmentation of OC and OD.
The encoder and decoder in this network are made up of dense blocks with a grouped convolutional layer in each block, allowing the network to acquire and convey spatial information from the image while simultaneously reducing the network's complexity.
To reduce spatial information loss, the optimal number of filters in all convolution layers were utilised.
In semantic segmentation, dice pixel classification is employed in the decoder to alleviate the problem of class imbalance.
The proposed network was evaluated on two publicly available datasets where it outperformed existing state-of-the-art methods in terms of accuracy and efficiency. 
For the diagnosis and analysis of glaucoma, this method can be used as a second opinion system to assist medical ophthalmologists.
\end{abstract}
\begin{IEEEkeywords}Optic Disc, Optic Cup, OC and OD Segmentation, Grouped Convolution,  Dense Network
\end{IEEEkeywords}

\section{\bf Introduction}
\sloppy
Glaucoma leads to permanent blindness worldwide, accounting for the second-highest number of instances of blindness after cataracts~\cite{tham2014global}. 
Glaucoma is a treatable condition that requires early detection to prevent progressive and irreversible loss of vision~\cite{wiggs2017genetics} and is typically classified into two types. One is closed-angle glaucoma and the second is open-angle glaucoma.
The first one is caused by an obstruction in the drainage angle due to parts of the iris, which results in the inability to drain the excess fluid and results in increased ocular pressure. 
Acute ocular pain, redness of the eye, sudden reduced vision, and elevated intra-ocular pressure are indications of closed-angle glaucoma~\cite{rizzo2017glaucoma}.
On the other hand, open-angle glaucoma is caused when the drainage angle is left open between the cornea and iris and has no early symptoms leading to difficulty in early diagnosis and treatment of this type of glaucoma~\cite{mowatt2008screening, naqvi2019automatic,khan2020region,tabassum2020cded,abdullah2021review,iqbal2022recent}.

The disease is diagnosed clinically by monitoring the pressure inside the eye over the course of many ophthalmologist examinations, where glaucoma causes the internal pressure to exceed 22 mmHg, when normal ocular pressure is between 12 and 22 mmHg~\cite{ohnell2019making}. 
Ophthalmoscopy (observe the health of optic nerve using dilation of pupil)~\cite{kelly2013teaching}, perimetry test (observe the change in the appearance of color or shape of the optic nerve)~\cite{mckendrick2005recent}, gonioscopy (identify changes in angle for drainage of fluid in eye)~\cite{sakata2008comparison}, and pachymetry (examine corneal thickness)~\cite{schroder2018comparison} are some of the main clinical tests used to diagnose glaucoma~\cite{thomas2011evaluation}.
All of the above-mentioned  techniques, however, are time-consuming, expensive, and may result in inter-observer variability~\cite{raghavendra2018deep}. 

Computer-based analysis of the retinal image, however, provides a reliable and effective method for detecting Glaucoma, allowing physicians and professionals to quickly evaluate each retinal image separately~\cite{hagiwara2018computer, tabassum2020cded, imtiaz2021screening}.
In the computerized scenario, the researchers primarily employ the Optic cup-to-disc ratio as an important sign as opposed to a set of tests performed in the clinical case for the detection of Glaucoma disease.
This is due to the fact that glaucoma is caused by the weakening of nerve fibers, which gradually affects the optic nerve, which is responsible for conveying signals from the retina to the brain.
Intraocular pressure (IOP), or the increase in fluid pressure within the inner region of the eye, damages the optic nerve.
Because the eye is unable to evacuate the surplus fluid, builds up pressure, causing the optic nerve fibers damage and thickening of the layers of retinal nerve fiber layers (RNFL).
The result is that the optic cup (OC) becomes larger than the optic disc (OD), a phenomenon named cupping.
Cupping or cup-to-disc ratio (CDR), which is the percentage of the OC's perpendicular diameter to the OD's perpendicular diameter, is an important metric for glaucoma diagnosis in computational approaches to the detection of glaucoma~\cite{hagiwara2018computer}.
The most efficient technique to identify glaucoma is to examine the optic nerve head (ONH), which is separated into two regions: the optic disc (OD) and the optic cup (OC)~\cite{naveed2017clinical,almazroa2015optic}. 

The traditional computer-aided techniques rely significantly on doctors' and ophthalmologists' knowledge, which results in calibrated measurements and observations based on the shapes and sizes of the optic disc and optic cup that have been recorded over time.
Optical nerve fibers of a healthy eye degenerate and widen the cup area as intraocular pressure (IOP) rises.
This results in an increase in CDR value, which is considered to be less than 0.6 for healthy eyes and more than 0.6 for eyes with glaucoma~\cite{ruengkitpinyo2015glaucoma}. 
However, relying solely on the cup-to-disc ratio to detect glaucoma is insufficient because some patients may have a large OC myopia~\cite{ruengkitpinyo2015glaucoma}. 
The ISNT rule, which stipulates that the rim thickness is thickest in the inferior area of a normal eye, followed by the nasal, temporal, and larger areas of the rim, is utilized for glaucoma screening since glaucoma causes OC size to rise, resulting in a violation of this criterion~\cite{mohamed2019automated}.
However, the exact segmentation of OC and OD is a critical task that is difficult in many ways.
For example, fundus cameras are employed to acquire retinal images, resulting in low resolution, poor contrast, noise, and artifacts in retinal images, introducing bright things (such as exudates) that may seem like high-intensity objects similar to OC and OD.
As a result, the required candidates, OC and OD, may be falsely detected.
Because of these difficulties, segmenting OC and OD is a difficult task. 
As a result, before segmenting the fundus image for blood vessels, the images must be pre-processed and unwanted noise and poor contrast must be removed~\cite{edupuganti2018automatic}. 
Due to the presence of blood vessels on the layer beneath the retina, detecting the OC borderline is difficult~\cite{soorya2018automated}.
Numerous authors have proposed levels of pre and post-processing, as well as segmentation, in order to address all of these flaws~\cite{al2018dense,qin2019optic}. 
Some studies pre-process the image to detect a region of interest (ROI) nearby OD before using a segmentation algorithm~\cite{al2018dense,qin2019optic}. 
These extra tasks necessitate expert input and additional time for feature extraction and processing, and accuracy varies from case to case, indicating a reliance on input data.

The use of neural networks, particularly state-of-the-art deep-learning-based computerized detection methods, not only eliminates the requirement for data pre-processing and post-processing \cite{khan2023simple, brahmavar2023ikd+,khan2023retinal, iqbal2023robust,qayyum2023semi,khan2022mkis,khan2023neural,arsalan2022prompt,iqbal2022g}. It also ensures competitive performance for combined OC and OD segmentation~\cite{fu2018disc,wang2019patch}. 
Typically, generative adversarial networks with a large number of parameters are used to achieve high accuracy for combined OC and OD segmentation~\cite{wang2019patch,jiang2019optic}.
However, the large number of parameters makes the inference slower and memory intensive \cite{khan2022t, khan2021rc,khan2020semantically}.
In \cite{fu2018joint}, the authors implemented an architecture with substantially less parameters while preserving the accuracy of the network, but performance is still dependent on the OD's precise localization.

In a previous study, a novel Cup Disc Encoder-Decoder Network (CDED-Net) design with reduced complexity for OC and OD segmentation from retinal fundus pictures was proposed~\cite{tabassum2020cded}.
The current study proposes a modified CDED-Net that removes superfluous processing blocks and parameters, reducing the complexity and total computational cost of the deep-learning model previously utilized to segment the cup and disc from retinal fundus images.
The new proposal is free from pre or post-processing due to the efficient encoder-decoder network design.
In addition, a more efficient encoder-decoder network is proposed to segment the OC and OD robustly.
To achieve this, fewer encoder layers are used, resulting in lower parameters and the preservation of semantic information.
On the decoder side, a dense layer is also added.
The major involvement of this work includes  
\begin{enumerate}
  \item As compared to existing state-of-the-art models, a unique encoder-decoder design is described that reduces computational complexity in the training and testing stages. The number of filters utilised for dense connection between the encoder and decoder sides is optimised in this architecture. Furthermore, this architecture efficiently removes  pre or post-processing steps.
  \item The use of grouped convolution in an intelligently designed encoder reduces the complexity of the design without impacting the performance of OC and OD segmentation. When compared to current state-of-the-art methods, this results in a more robust evaluation of glaucoma with less system complexity.
\end{enumerate}

The layout of the paper is given below.
The relevant literature is discussed in Section 2.
The approach and network structure for OC and OD segmentation are discussed in Sections 3 and 4, respectively.
Section 5 contains the experimental data as well as a comprehensive comparison with state-of-the-art methods, with Section 6 providing a full justification for the proposed method.
Lastly, section 7 concludes the argument and brings the research project to a culmination.

\section{Related Work}
Segmenting the OC and OD in order to determine the precise cup-to-disc ratio is a key step in glaucoma diagnosis.
Thresholding, region-based methods, and active contour-based methods are among the OC and OD segmentation methods described in this literature review, which are broadly classified as deep learning-based and image-processing techniques.
Thresholding is the primary technique for segmenting a binary image, in which the segmented image is obtained by applying a threshold value \cite{khan2022width,khawaja2019multi,khawaja2019improved}.
Various thresholding algorithms, such as adaptive thresholding, have been employed for the segmentation of OC and OD due to their efficacy and ease of implementation~\cite{agarwal2015novel}. 
Otsu thresholding is a prominent technique to segment OD and OC that uses the red channel of the RGB picture for OD segmentation and the green channel of the ROI image for OC segmentation.
By extracting and adding features, quality-independent segmentation, and intensity value can be computed using a threshold.
Another method for segmentation is based on the median, mean, and Otsu thresholding.
However, the problem with these methods is that patients' retinal color varies, reducing the effectiveness of such solutions.

The use of an active contour model~\cite{kumar2015active} is also suggested for OD segmentation.
The initialization of the contour model improves the efficiency of such a system.
The existence of noise and pathologies in an image, on the other hand, causes this system to become stuck on a local minimum~\cite{thakur2018survey}.
Similarly, in~\cite{haleem2018novel}, classification errors were reduced by utilizing an adaptive deformable model that recognizes shape changes and irregularities.
Furthermore, utilizing a superpixel classification-based approach~\cite{cheng2013superpixel}, the segmentation of the cup and disc has been considered.
The disadvantage of this method is that fluctuations in cup size lead to inaccurate cup estimation.
To segment OC and OD simple linear iterative clustering (SLIC) is used in the superpixel-based approach  for the extraction of superpixels from the image and a classifier is used to distinguish between background, OC, and OD regions.

In~\cite{soorya2018automated}, the use of intensity-based thresholding and geometric features were recommended for OD segmentation and removal of noisy pixels.
In~\cite{thakur2018survey} a survey is presented in which a technique based on vessel tracking bending was used to track and observe abrupt alterations in blood vessels, however, thresholding-based methods were not efficient in low-contrast images.
A method was suggested in~\cite{cheng2011automatic} to reduce peripapillary atrophy by combining several techniques to obtain an accurate disc boundary.
Some studies employed region-based segmentation methods because regional data is more resistant to changes in contrast and intensity.
CAR and CDR were estimated in~\cite{joshi2011optic} to develop an upgraded segmentation technique for OC and OD.
Furthermore, the histogram of the image's red component was used in~\cite{issac2015adaptive} to choose a threshold for OD segmentation from the green component.

Despite their utility, these traditional approaches suffer from a variety of artifacts, inadequate lighting, and noise in fundus images.
To address these challenges, several pre-processing techniques are performed to produce high-quality retinal pictures, which improves segmentation results significantly.
As a result, pre-processing techniques, which aim to remove noise, artifacts, and difficulties with non-uniform lighting in fundus images, have become an important part of the OC and OD segmentation pipeline.

Robust segmentation is achieved using machine learning approaches but at the cost of high computational complexity. 
In~\cite{septiarini2018automatic}, a glaucoma detection technique is proposed, in which image classification utilizing statistical characteristics and K-nearest neighbor is employed.
Similarly, automated regression is said to be used to determine the proper OC and OD boundary~\cite{sedai2016segmentation}.
An effective and accurate approach of OD localization in images with noise and other lesions was proposed in~\cite{akyol2016automatic}.
Deep learning-based methods, in which complicated features are automatically learned through training, have recently become popular.
In~\cite{sevastopolsky2017optic}, a transformation of the original U-Net CNN was provided to segment OC and OD, where the image dimensions of the given image are expanded by going through the contracting and expansive network paths with upsampling layers.
This approach produces high-quality OC and OD segmentation with the shortest forecast time.
In~\cite{zilly2017glaucoma}, an ensemble learning architecture for OC and OD segmentation was inspired by a CNN.
To design convolutional filter learning architecture, the most relevant points were determined using the entropy sampling method.
This approach is appropriate for tiny datasets.
The authors of~\cite{edupuganti2018automatic} proposed a fully convolutional neural network (FCNN) with upsampling layers, a VGG-16 encoder, and a decoder to achieve the segmented image.

\section{Proposed Methodology}
In this paper, we propose the EDDense-Net architecture for efficient OC and OD segmentation from fundus images.
Traditional deep learning techniques, which use several convolutional layers followed by pooling operations and may lose local information about some essential pixels, are targeted by the proposed architecture.
As a result, detecting OC and OD accurately has become difficult by using traditional deep learning techniques.
The aim of this study is to propose a network that maintains a fine balance between network complexity and performance.

\subsection{Overview of Proposed Architecture}
The general architecture of the suggested technique is depicted in Figure~\ref{fig1}, which depicts the configuration of the encoder-decoder block's deep feature concatenation.
EDDense-Net receives the original image as input without using any pre-processing technique in this design.
The suggested EDDense-Net makes use of the deep feature to import and aggregate high-frequency information from the relevant layers.
With the use of pixel-wise segmentation, EDDense-Net can also recognize vessels from low-quality images and non-uniform light.
The proposed method produces a binary image with a representation of `1' for the vessel area as output and `0' for the background.

\subsection{OC and OD Segmentation using EDDense-Net}
EDDense-Net is designed to tackle those semantic segmentation issues that are not resolved by traditional deep learning techniques by addressing dense connectivity with the following three key design principles: 
\begin{enumerate}
    \item Spatial information loss is reduced by using lesser convolutional and pooling layers.
    \item Immediate spatial information transfer between the layers is provided by the network using dense concatenation within the dense block. 
    \item The faster convergence of the network is achieved by transferring edge information from the first layers of the encoder to the last layer of the decoder. 
\end{enumerate} 
The encoder is made up of five dense blocks, each with a group convolutional layer.
The decoder has a structure that is similar to the encoder.
In most existing systems, basic convolution is utilized with dense layers.
The process of utilizing a varying number of convolution filter groups on the same image is known as grouped convolution.
The number of kernels in each layer is raised in grouped convolution to achieve more than one convolution pathway and improve performance.
Grouped convolution increases the network's expressive capacity without increasing the number of parameters.
In the network~\cite{cohen2016group}, layers of grouped convolution can be simply created with relatively minimal computational overheads.
With grouped convolution, large networks may be created simply by replicating the integrated block of the filter group as many times as desired.
Each filter only convolves on a subset of the feature maps obtained by the filter group, minimizing the number of computations required to obtain output feature maps.
In the last layer of the suggested architecture, dice pixel classification was used instead of simple pixel classification, which leverages generalized dice loss to overcome the class imbalance problem in semantic segmentation.

\subsection{EDDense-Net Encoder}
\label{encoder}
As demonstrated in Figure~\ref{fig1}, EDDense-Net is a densely connected fully convolutional network that uses a total of 5 dense blocks for both encoder and decoder.
The encoder is made up of five dense blocks, each of which has a convolutional layer and a grouped convolutional layer.
Each encoder dense block begins with a convolutional layer and concludes with a pooling layer that reduces the feature map's size.
As a result, the EDDense-Net encoder applies a continual convolutional operation to the image and the feature map, which then flows through the network in a feed-forward method until the image is represented by small features.
The issue with CNN is that the max-pooling operation (after convolution) results in the loss of spatial information.
The loss of relevant information is covered by grouped convolution with dense layers in EDDense-Net.
The encoder is made up of five dense blocks with five grouped convolution layers and five max-Pooling layers in the proposed architecture.
For a $640 \times 640$ input image, the final feature map is $20 \times 20$.
Table~\ref{Encoder} defines the feature empowerment inside each encoder dense block and demonstrates how the bottleneck layer reduces the depth of the feature map.
It also illustrates how the EDDense-Net encoder structure in terms of convolution and pooling.
The table also includes information on the number of parameters in the encoder's layers. 

\subsection{EDDense-Net Decoder}
\label{decoder}
\begin{figure*}[t]
\centering
  \includegraphics[scale=0.10]{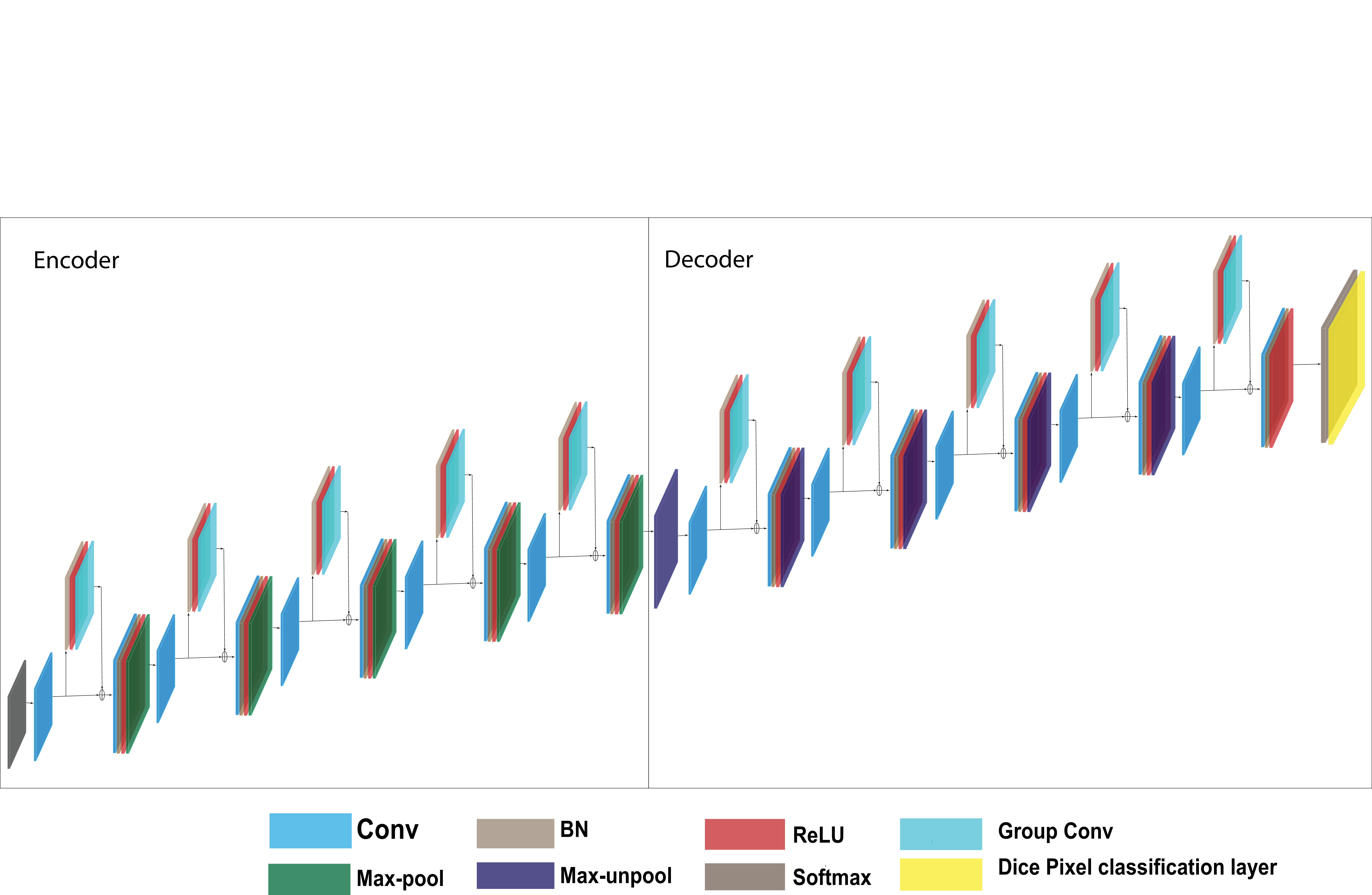}
  \caption{Proposed Fully Dense Encoder Decoder Network architecture.}
  \label{fig1}
   \vspace{0.1cm}
\end{figure*}

As illustrated in Figure~\ref{fig1}, the decoder in EDDense-Net uses the reverse procedure of the encoder, with each dense block starting with a Max-Unpool layer that is responsible for gradually increasing the size of the feature map.
A convolution layer is applied after each unpooling layer, followed by grouped convolution.
The outer dense routes begin at the encoder dense block's first convolutional layer and end at the concatenation layer of each decoder dense block.
To reduce latency, these outer dense paths provide immediate edge information from encoder to decoder.
The EDDense-Net decoder, in particular, receives a $20 \times 20$-pixel input from the encoder and outputs a final feature map of the same size as the input image.
The network's complexity is reduced by using grouped convolution in each decoder block.
The ReLU and Dice pixel classification layer is used to overcome the class imbalance problem in semantic segmentation at the end of the last layer in the decoder (5th decoder dense block).
The layer-wise feature map details for the proposed decoder are shown in Table~\ref{Decoder}.

\section{\bf \textsc{Databases and Performance Metrics}}
\label{sec:Databases_Methodlogy}
The proposed model and the databases used for its assessment are discussed in this section. 

\subsection{Retinal Fundus Image Databases}
For performance evaluation of our developed model for the tasks of semantic segmentation (pixel-wise) of both OD and OC, we used two publicly available retinal image databases: DRISHTI-GS~\cite{sivaswamy2015comprehensive} and RIMONE~\cite{fumero2011rim}.

\subsubsection{Drishti-GS}
This database comprises 101 retinal images taken at a field of view of 30 degrees with a centered-optical disc and dilated pupils.
An expert from Aravind Eye Clinic (Madurai, India) annotated each of these collected images.
All of the images were saved in PNG format with no compression and have a common resolution of $2896 \times 1944$.
In all images of the DRISHTI-GS1, the average OD and OC borders are shown, which are based on four expert hand labelings.

\subsubsection{RIM-ONE}
Researchers can use this retinal dataset, which contains 159 images, to evaluate OD/OC segmentation models.
74 of the 159 retinal images belong to glaucoma-affected eyes, whereas the remaining 85 belong to healthy people.
These photographs were taken at three Spanish hospitals and graded by two experts.

\begin{table*}[httbp]
\centering
\caption{The layer-wise feature map details for the downsampling block of the proposed model}
\scalebox{1}{
		\resizebox{1\textwidth}{!}{
			\setlength\extrarowheight{3pt}
			
\begin{tabular}{lllll}
\hline
\textbf{Layer}              & \textbf{Name and Size}  & \textbf{Filters size} & \textbf{The size of the feature map} & \textbf{Parameter information} \\ \hline
The 1st block of convolution & Conv1\_1/3x3x3  & 32                     & 640x640x32                       & 896                      \\
                            & groupedconv\_1/3x3x1 & 32                        & 640x640x32               &320
                            \\
                            \\
1st Pooling layer                   & Pool1/2x2         & 2x2                     & 320x320x32          & 0                        \\
The 2nd block of convolution & Conv2\_1/3x3x32  & 64                     & 320x320x64               &  18496                     \\
                        
                            & groupedconv\_2/3x3x1 & 64                       & 320x320x64               &640
                            \\
                            \\
2nd Pooling layer                   & Pool2\_1/2x2         & 2x2                   & 160x160x64      & 0                          \\
The 3rd block of convolution & Conv3\_1\_1/3x3x64  & 128                     & 160x160x128         &73856                       \\
                            
                            & groupedconv\_3/3x3x1 & 128                       & 160x160x128            &1280
                            \\
                            \\
The 3rd Pooling layer                   & Pool$2_2/2x2 $        & 2x2            & 80x80x128               &0                                \\
The 4th block of convolution  & Conv3$\_1\_2/3x3x128$  & 128                     & 80x80x128              &147584                       \\
                            
                            & groupedconv$\_4/3x3x1$ & 128                       & 80x80x128                 &1280
                             \\
                             \\
                            
The 4th Pooling layer              & Pool$2_3/2x2 $        & 2x2            & 40x40x128                  &0
 \\
 The 5th block of convolution  & Conv3$\_1\_3/3x3x128$  & 128                     & 40x40x128              &147584                       \\
                            
                            & groupedconv$\_5/3x3x1$ & 128                       & 80x80x128                 &1280
                            \\
                            \\
 The 5th Pooling layer                & Pool3/2x2         & 2x2            & 20x20x128                      &0                                      \\ \hline
\end{tabular}
}}
\label{Encoder}
\end{table*}

\begin{table*}[httbp]
\centering
\caption{The layer-wise feature map details for the upsampling block of the proposed model}
\scalebox{1}{
		\resizebox{1\textwidth}{!}{
			\setlength\extrarowheight{3pt}
\begin{tabular}{lllll}
\hline
\textbf{Layer}              & \textbf{Name and Size}  & \textbf{Filter size} & \textbf{The size of the feature map} & \textbf{Parameter information} \\ \hline
5th unpooling layer                & decoder3\_unpool             &                         & 40x40x128     &0                           \\
5th block of convolution & decoder3\_conv2\_1\_3/3x3x128  & 128                     & 40x40x128     &147584                   \\
                        
                        & groupedconv\_6/3x3x1 & 128                       & 40x40x256               &1280
                    
                            \\
                            \\
4th unpooling layer                & decoder2\_unpool\_3        &      & 80x80x128                   &0  
\\          
4th block of convolution &  decoder3\_conv2\_1\_2/3x3x128  & 128            & 80x80x128           &147584
\\
                        
                        & groupedconv\_7/3x3x1 & 128                       & 80x80x128                 &1280
                        \\ 
                        \\

3rd unpooling layer                & decoder2\_unpool\_2         &      & 160x160x128                   &   0                        \\
3rd block of convolution & decoder3\_conv2\_1\_1/3x3x128  & 128                     & 160x160x128      &  147584                 \\
                        
                        & groupedconv\_8/3x3x1 & 128                       & 160x160x128                 & 1280
                        \\             
                        \\
2nd unpooling layer                 & decoder2\_unpool\_1       &        & 320x320x64                  &0                           \\
2nd block of convolution & decoder3\_conv2/3x3x64  & 64                    & 320x320x64      & 36928                  \\
                        
                        & groupedconv\_9/3x3x1 & 64                       & 320x320x64                  &640
                        \\             
                        \\
1st unpooling                 & decoder1\_unpool         &      & 640x640x32                  & 0                          \\
1st block of convolution & decoder1\_conv2/3x3x32  & 32                   & 320x320x64      & 9248                  \\
                       
                        & groupedconv\_10/3x3x1 & 32                      & 320x320x64                  & 320                 \\ \hline
\end{tabular}
}}
\label{Decoder}
\end{table*}

\subsection{Performance Metrics}
We have used standard evaluation metrics for assessing the performance of our developed model on the publicly available datasets of Drishti-GS and RIM-ONE. 
We have used the evaluation metrics including dice coefficient (F1 score), Jaccard (O), specificity, sensitivity, overlapping error (E) and balanced accuracy (BA). 
We aim to evaluate our developed EDDense\_Net model for the OC/OD segmentation compared to the graded ground truth from experts. 
The selected evaluation metrics are as given below:

\begin{equation}
Dice\_Coefficient (DC) = \frac {TP + TP}{TP + TP + FP + FN}
\end{equation}

\begin{equation}
Jaccard(O) = \frac {TP}{FN + FP + TP}
\end{equation}

\begin{equation}
E= 1-\frac {Area(S\cap G)}{Area(S\cup G)}
\label{Error}
\end{equation} 

\begin{equation}
Sen = \frac {TP}{FN + TP}
\end{equation}

\begin{equation}
Sp = \frac {TN}{FP + TN}
\end{equation}

\begin{equation}
BA=\frac {Sen + Sp}{2}
\label{BAcc}
\end{equation}

False positive, true positive, false negative and true negative are abbreviated as FP, TP, FN, and TN respectively. 
Also, the specificity and sensitivity are referred to by $Sp$ and $Sen$ respectively. 

\section{\bf \textsc{Results}}
\label{sec:Results}
We present our experimental results both graphically and in tabular form in this section. 
Furthermore, the details of the databases and the evaluation metrics used are also unfolded. 

\subsection{Performance Metrics}
We evaluated our developed model for the segmentation  using two databases i.e. Drishti-GS and RIM-ONE and presented the results in Table~\ref{OD_OC_Drishti} and Table~\ref{OD_OC_RIMONE} respectively. 
Table~\ref{OD_OC_Drishti} indicates that the proposed model achieved the highest dice coefficient (F1 score), Jaccard coefficient (O), and sensitivity for OC detection and comparable results for OD detection on test images from the Drishti-GS dataset. 
Table~\ref{OD_OC_RIMONE} shows that the proposed model achieved the highest dice coefficient (F1 score) and sensitivity while the second highest Jaccard coefficient (O) for OC detection on test images from the RIM-ONE dataset. 
It is also evident from this table that the proposed model achieved the highest dice coefficient (F1 score) and sensitivity while the second highest Jaccard coefficient (O) for OD detection on test images from the RIM-ONE dataset. 
We observed that the proposed EDDense-Net performed much better for OC segmentation than the state-of-the-art methods. 
For OD segmentation, the proposed EDDense-Net performance is comparable with the state-of-the-art methods. 
These results proved that the proposed model is robust and reliable and advocates for its use for glaucoma diagnosis.

\subsection{Comparison with state-of-the-art}
\begin{figure*}
  \centering
  \resizebox{0.78\textwidth}{!}{%
  \begin{tabular}{cccc}
       \includegraphics[width=0.14\textwidth]{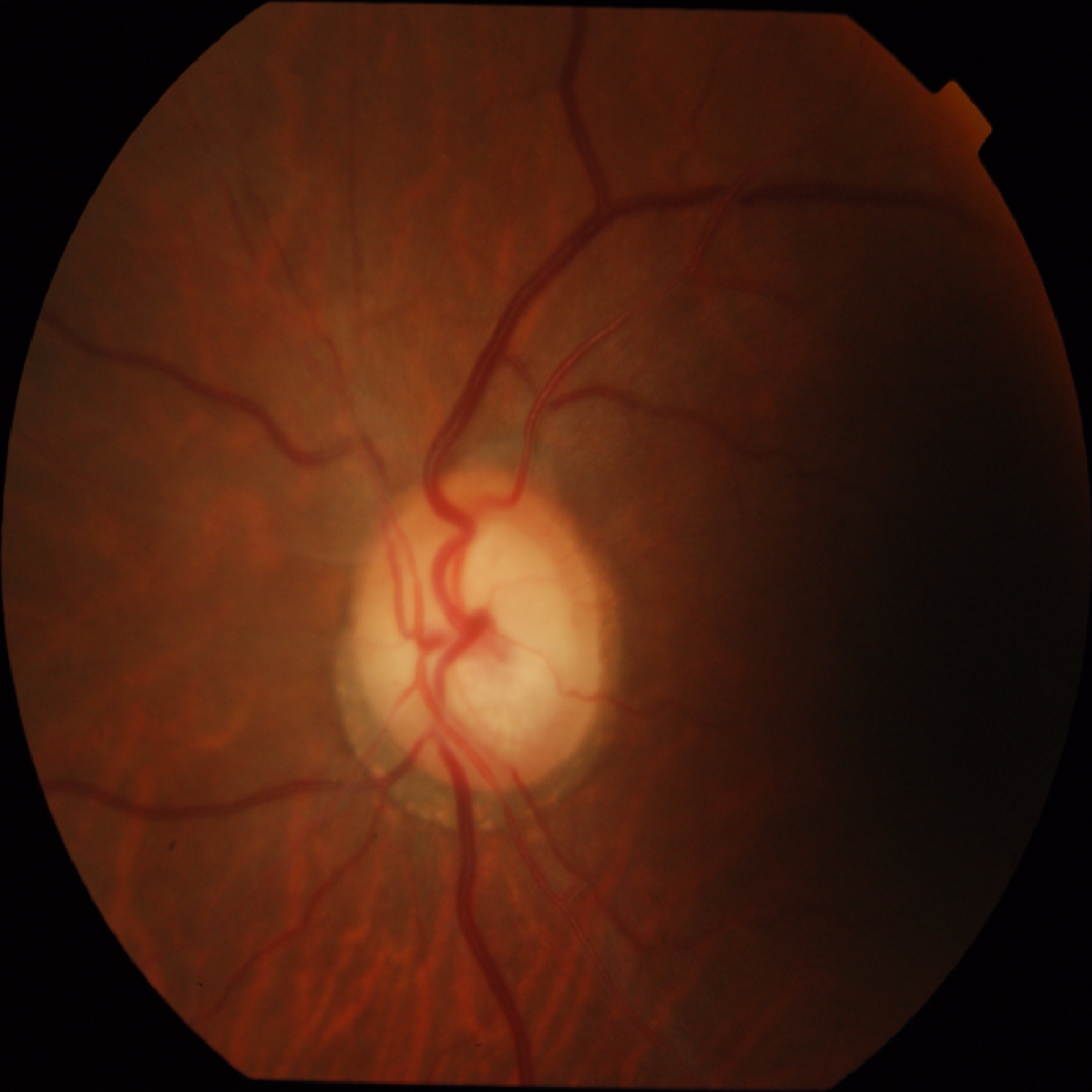}  & 
       \includegraphics[width=0.14\textwidth]{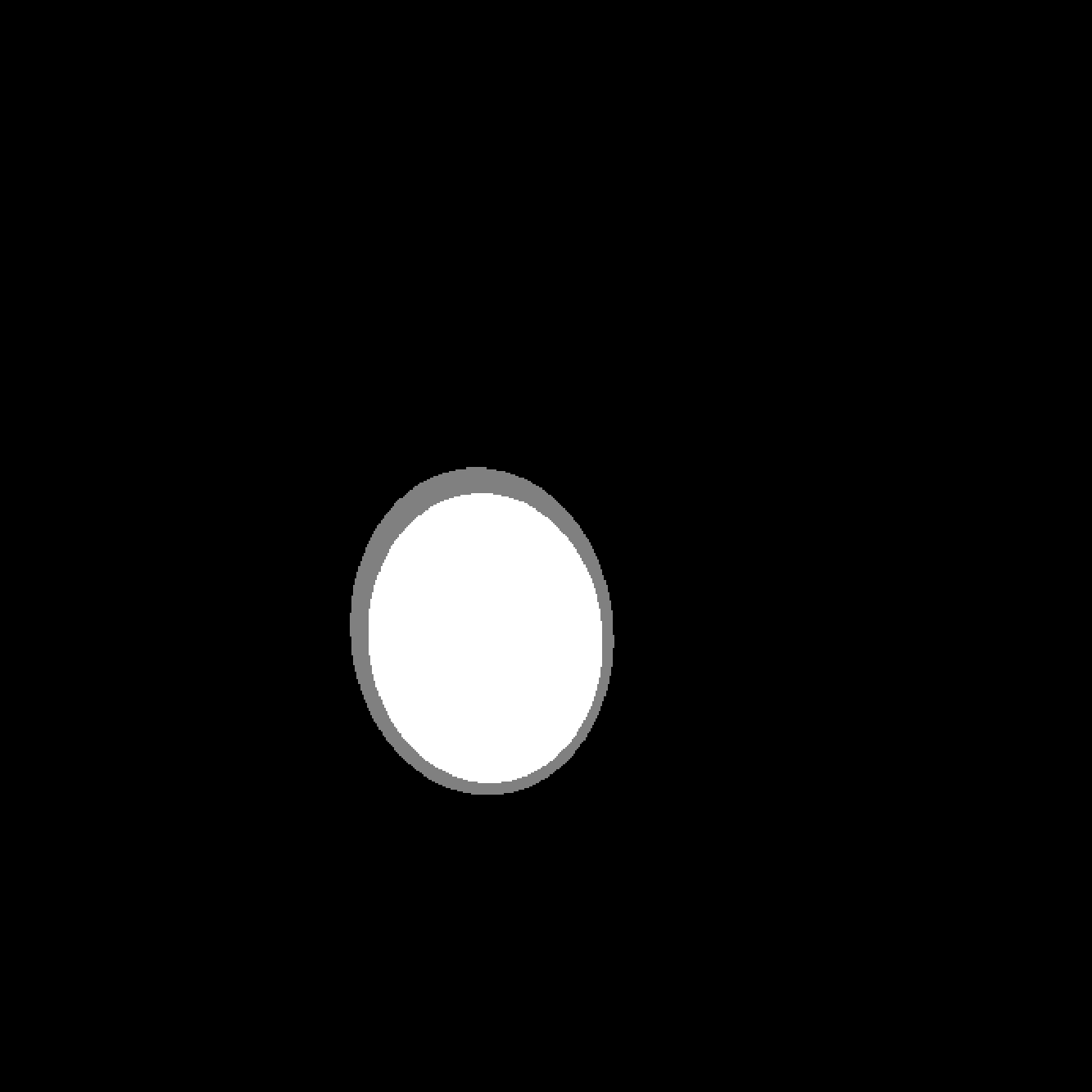}    & 
       \includegraphics[width=0.14\textwidth]{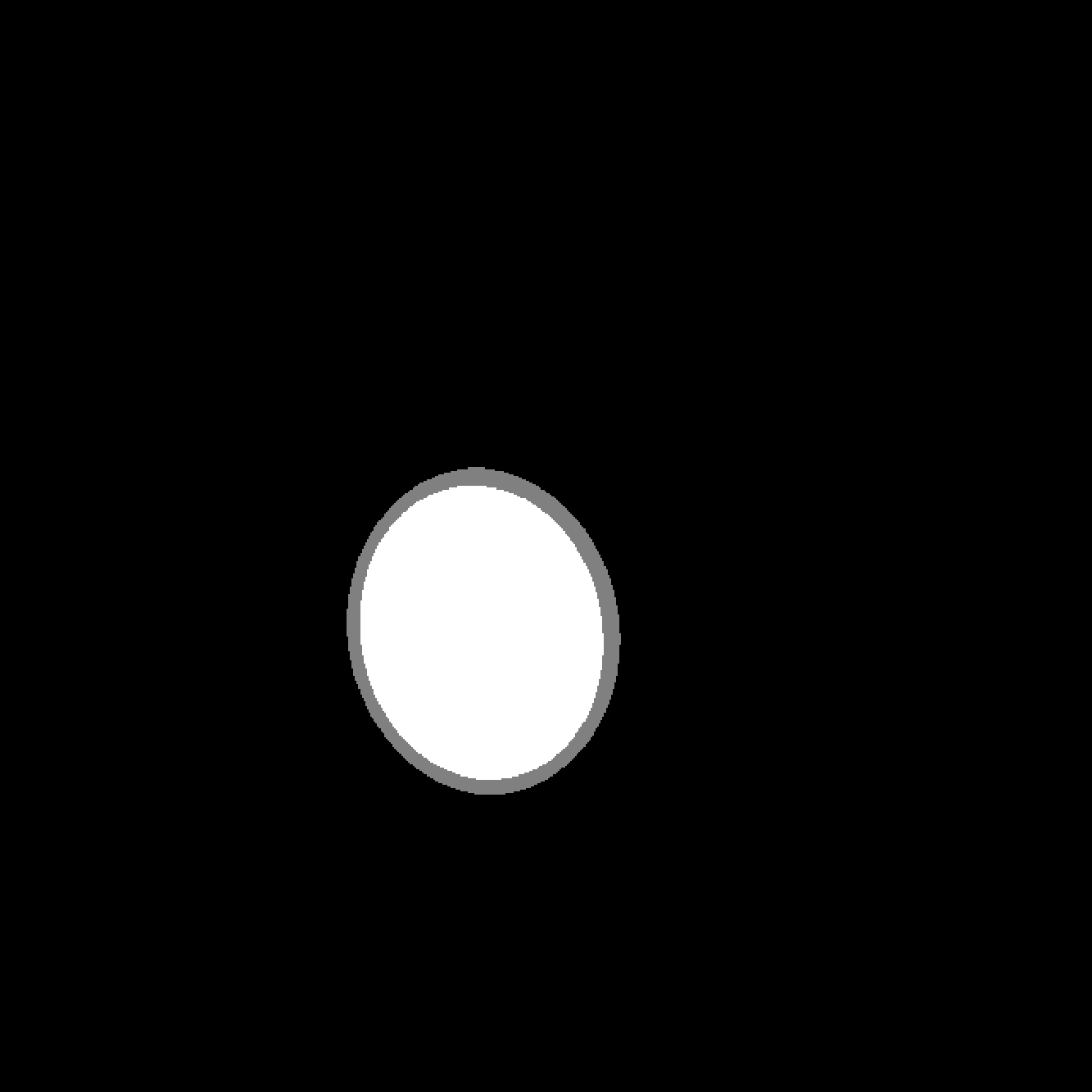}  \\
       \includegraphics[width=0.14\textwidth]{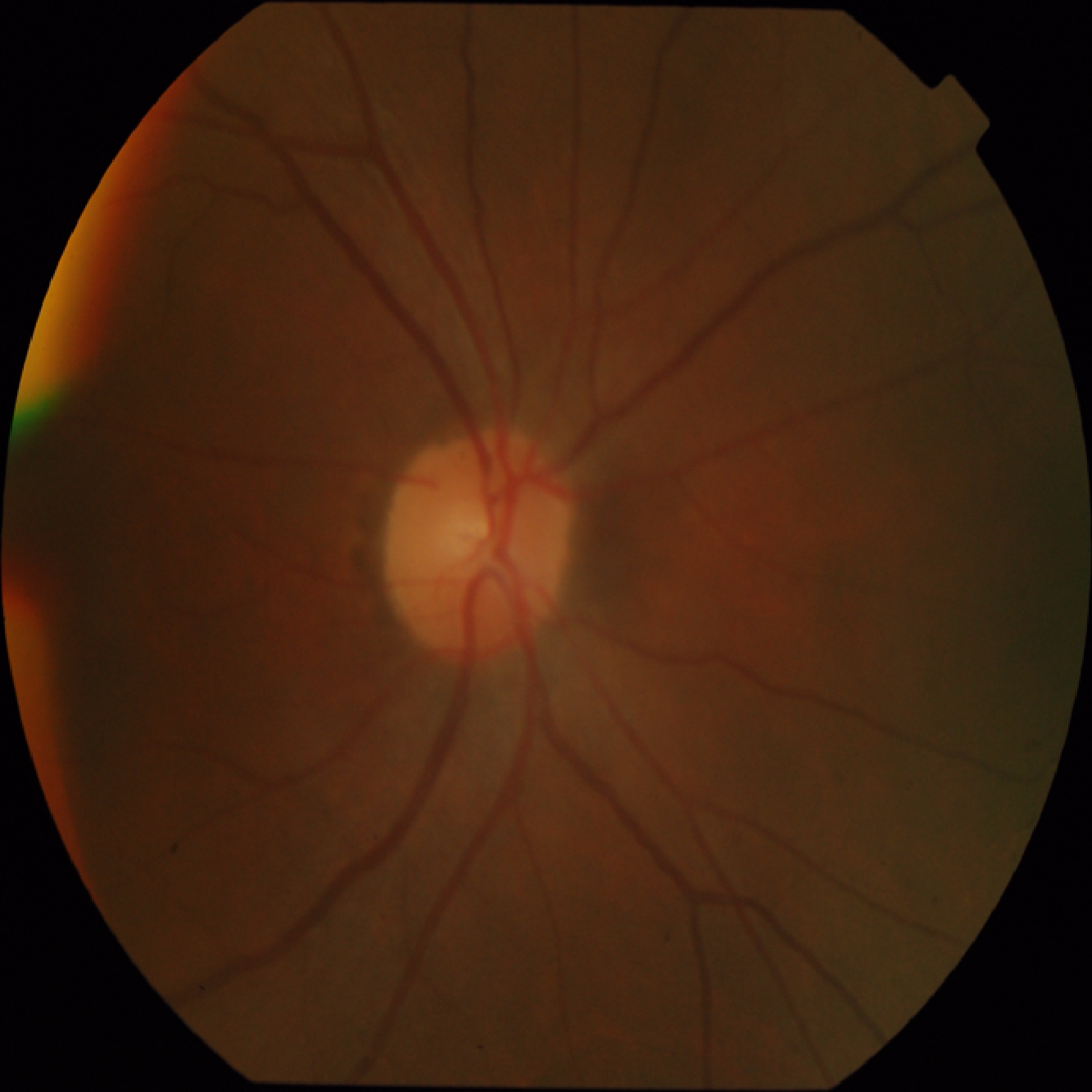}  & 
       \includegraphics[width=0.14\textwidth]{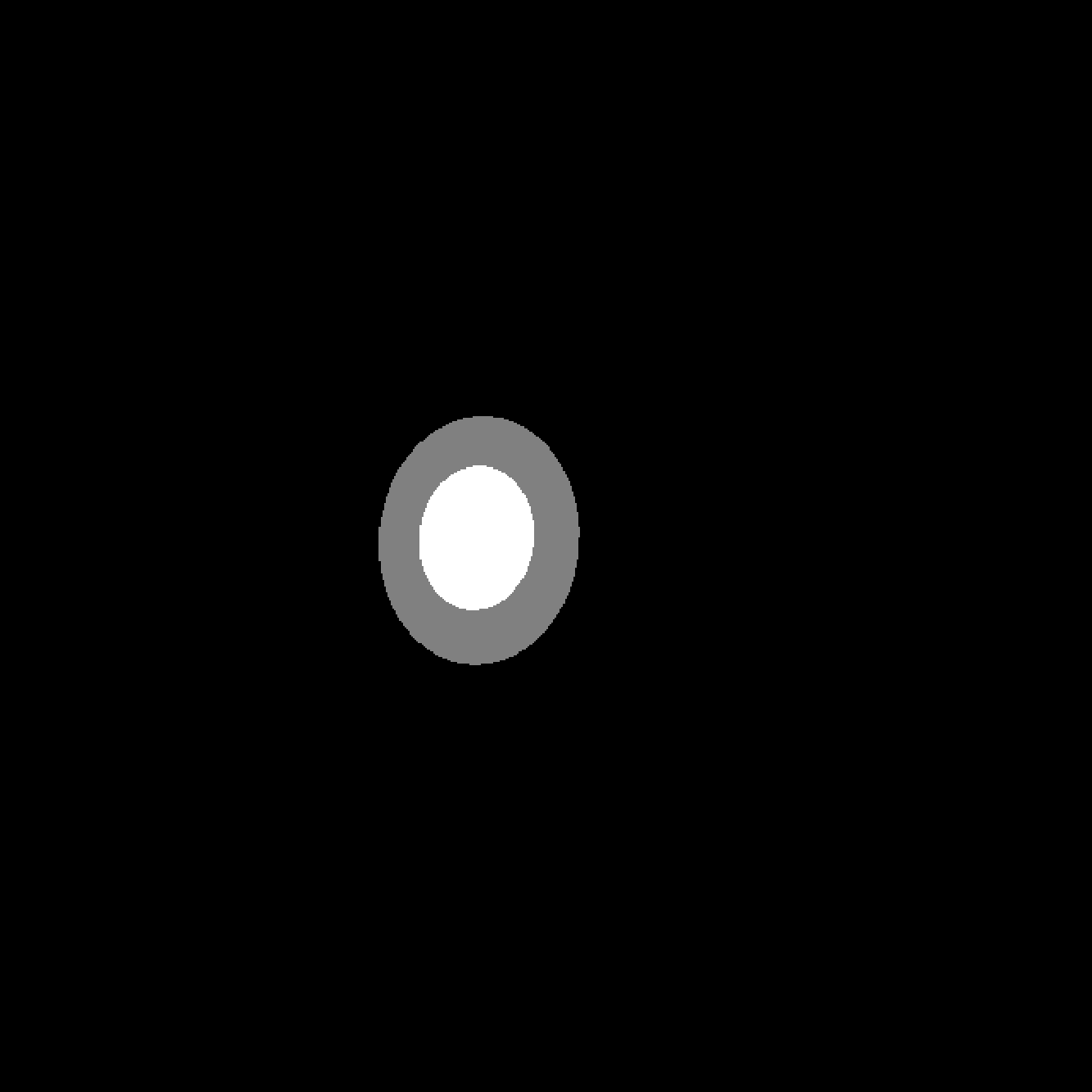}    & 
       \includegraphics[width=0.14\textwidth]{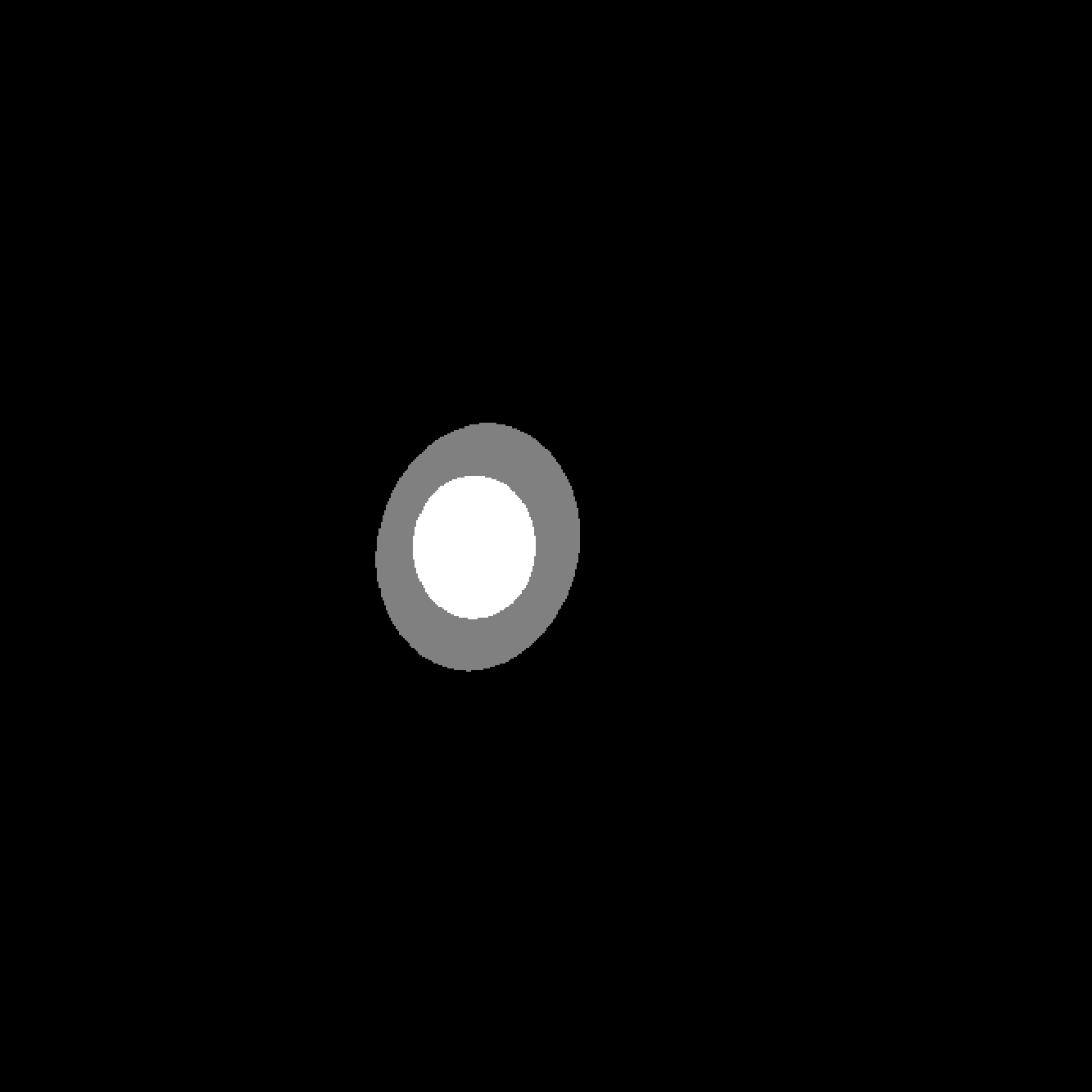}  \\       
       \includegraphics[width=0.14\textwidth]{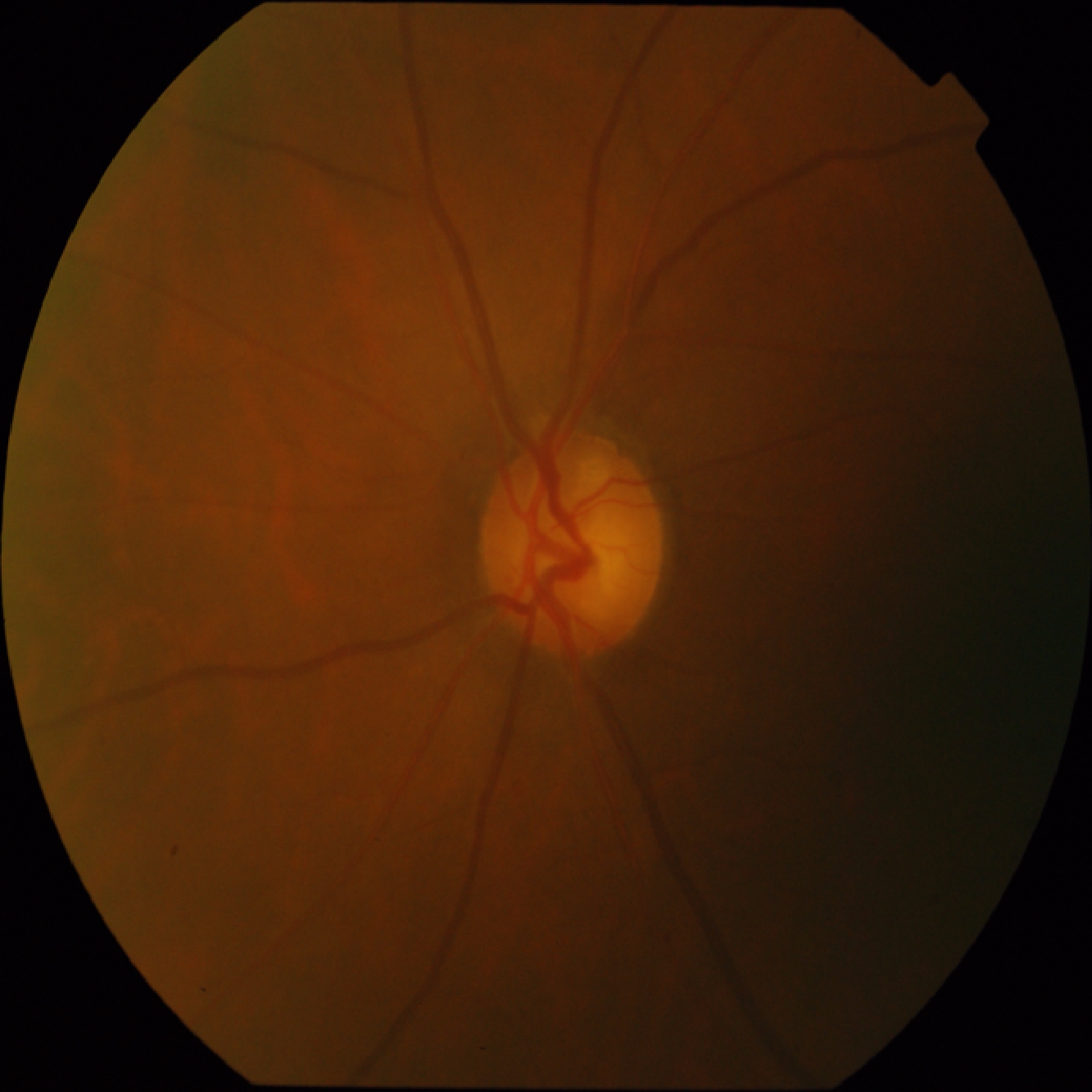}  & 
       \includegraphics[width=0.14\textwidth]{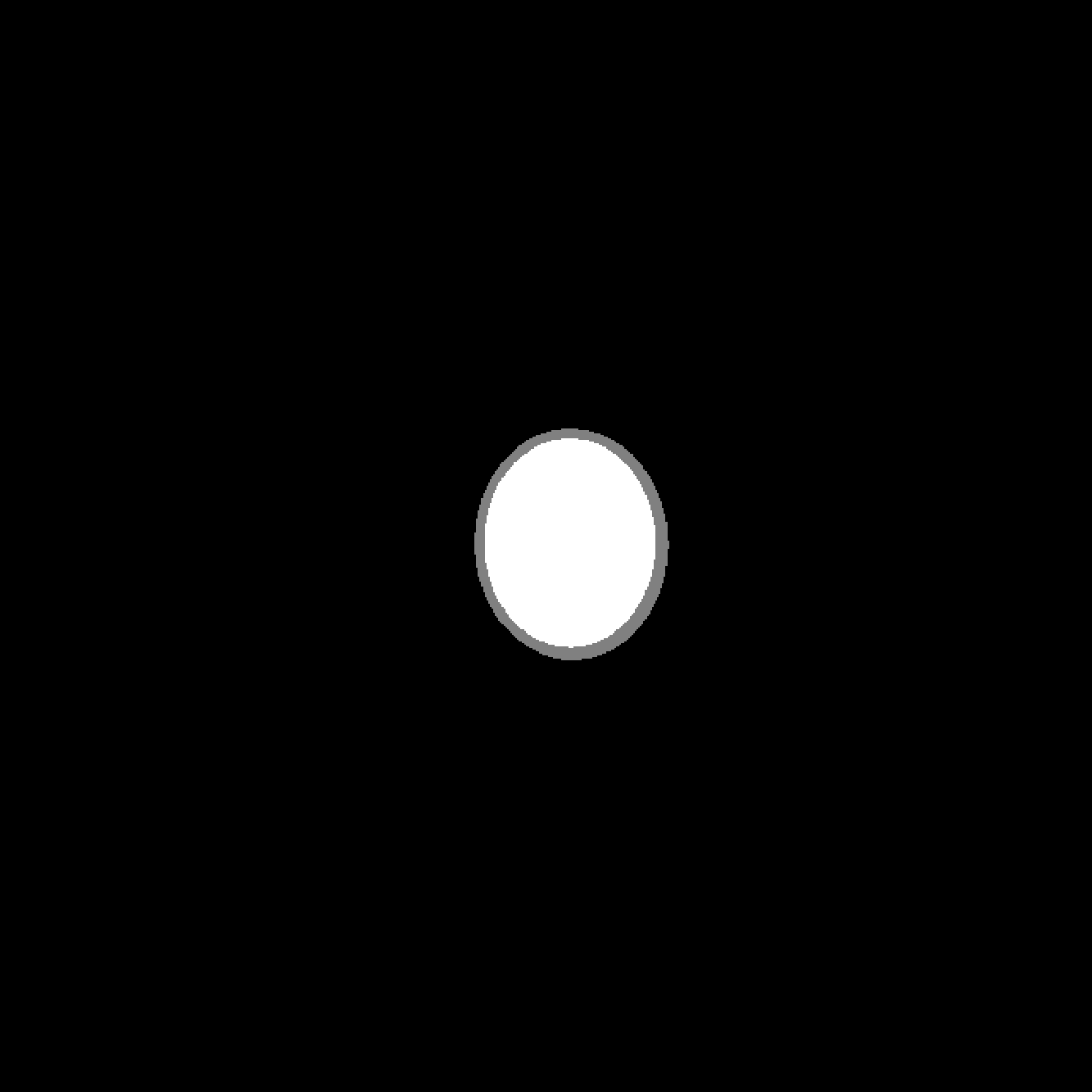}    & 
       \includegraphics[width=0.14\textwidth]{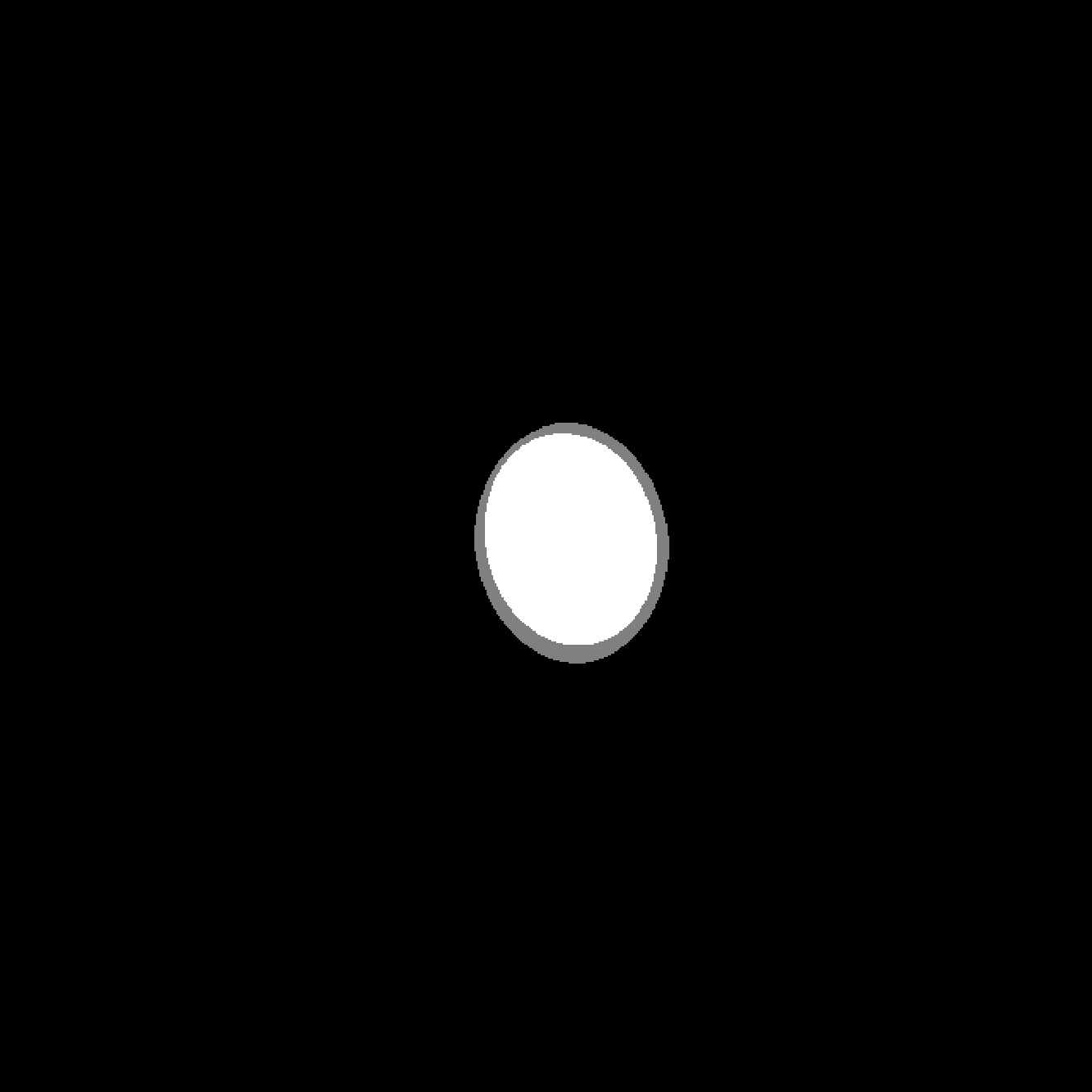}  \\       
       \includegraphics[width=0.14\textwidth]{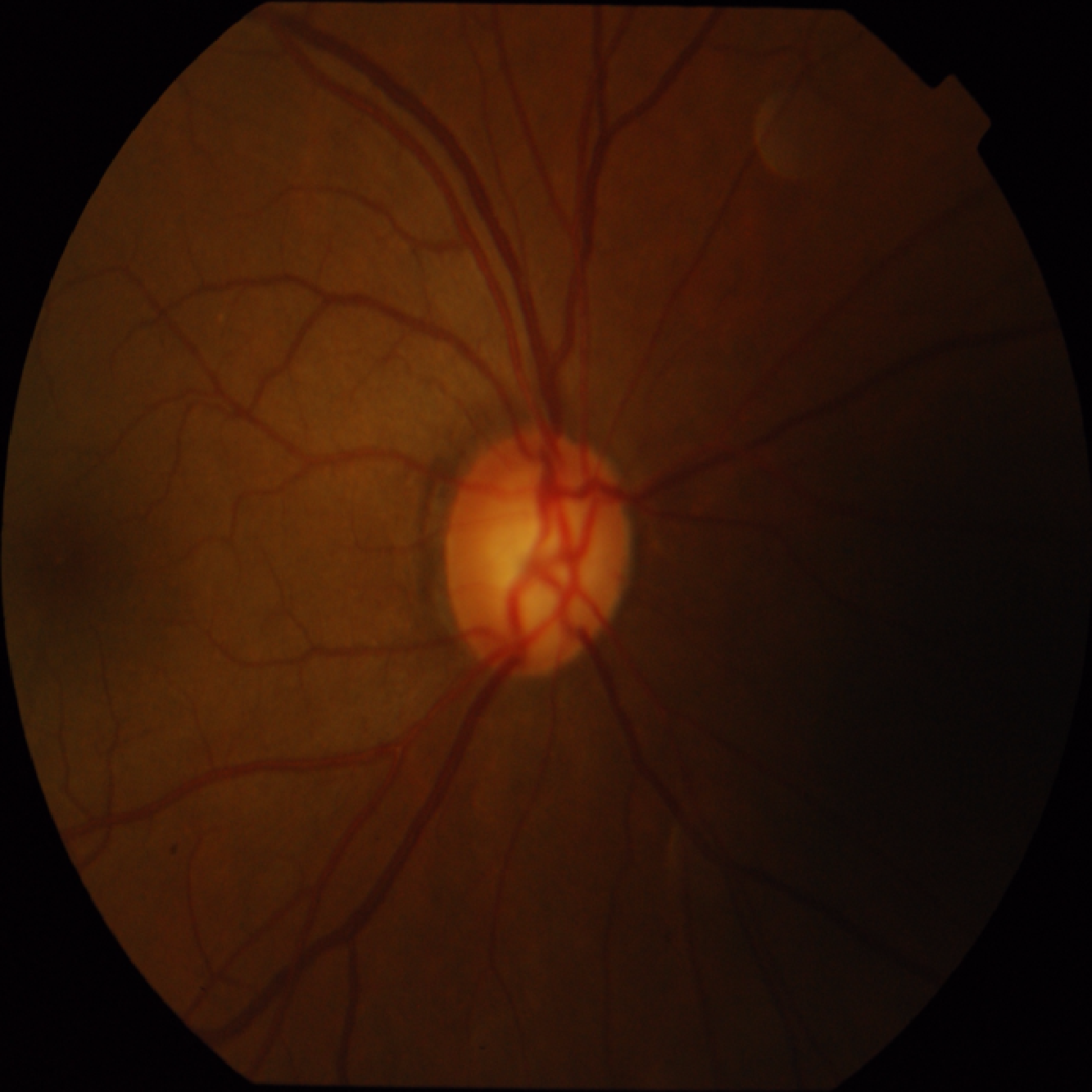}  & 
       \includegraphics[width=0.14\textwidth]{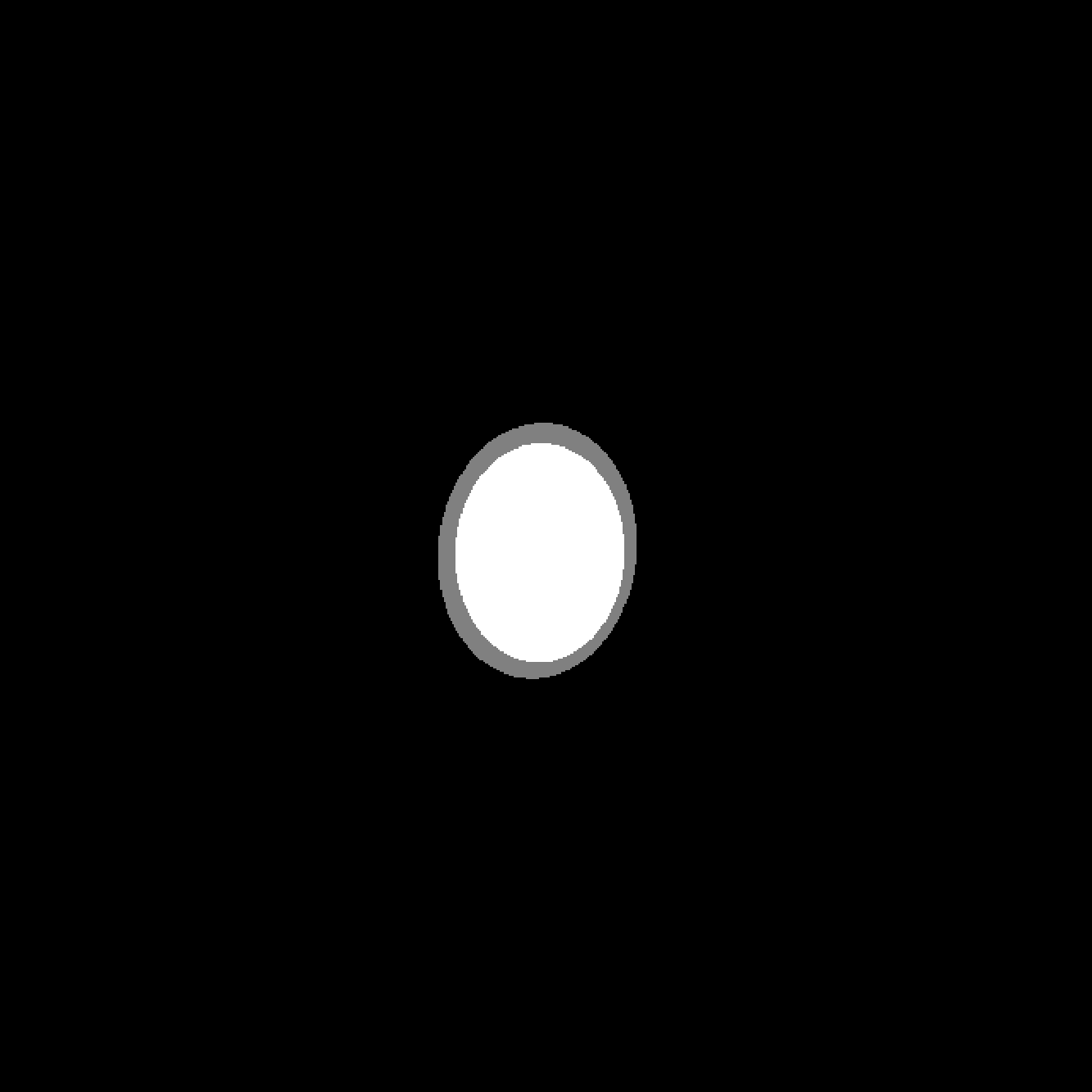}    & 
       \includegraphics[width=0.14\textwidth]{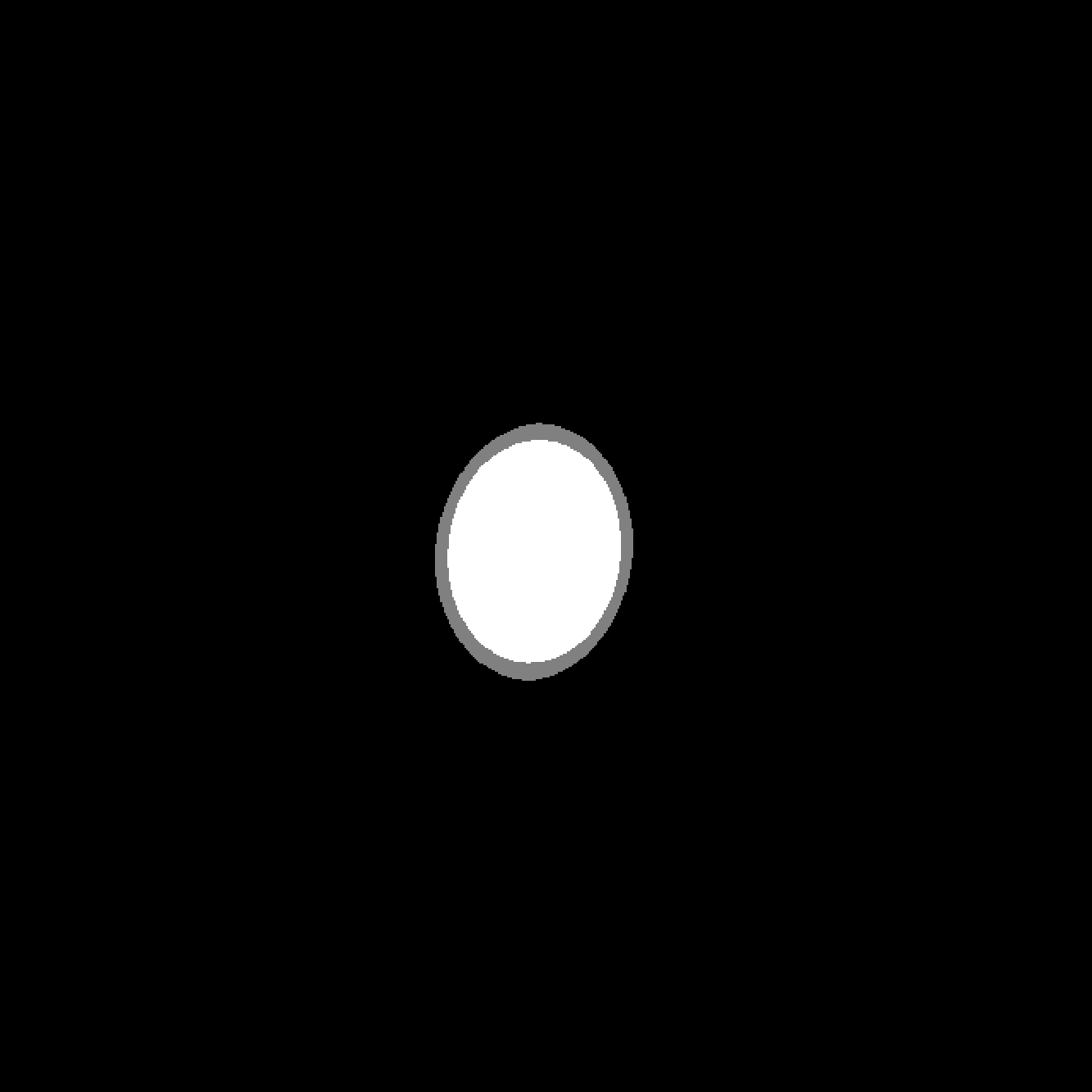}  \\       
       \includegraphics[width=0.14\textwidth]{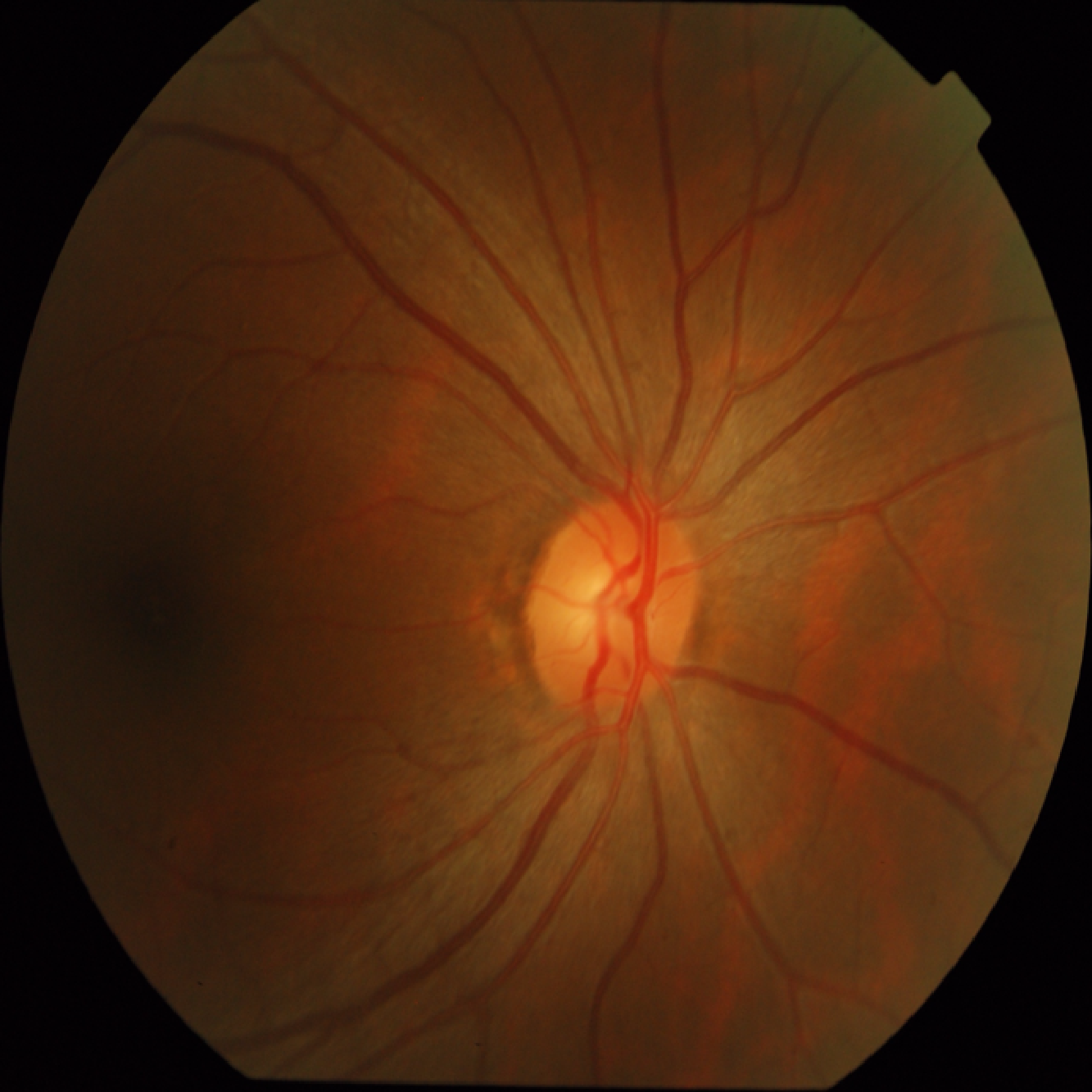}  & 
       \includegraphics[width=0.14\textwidth]{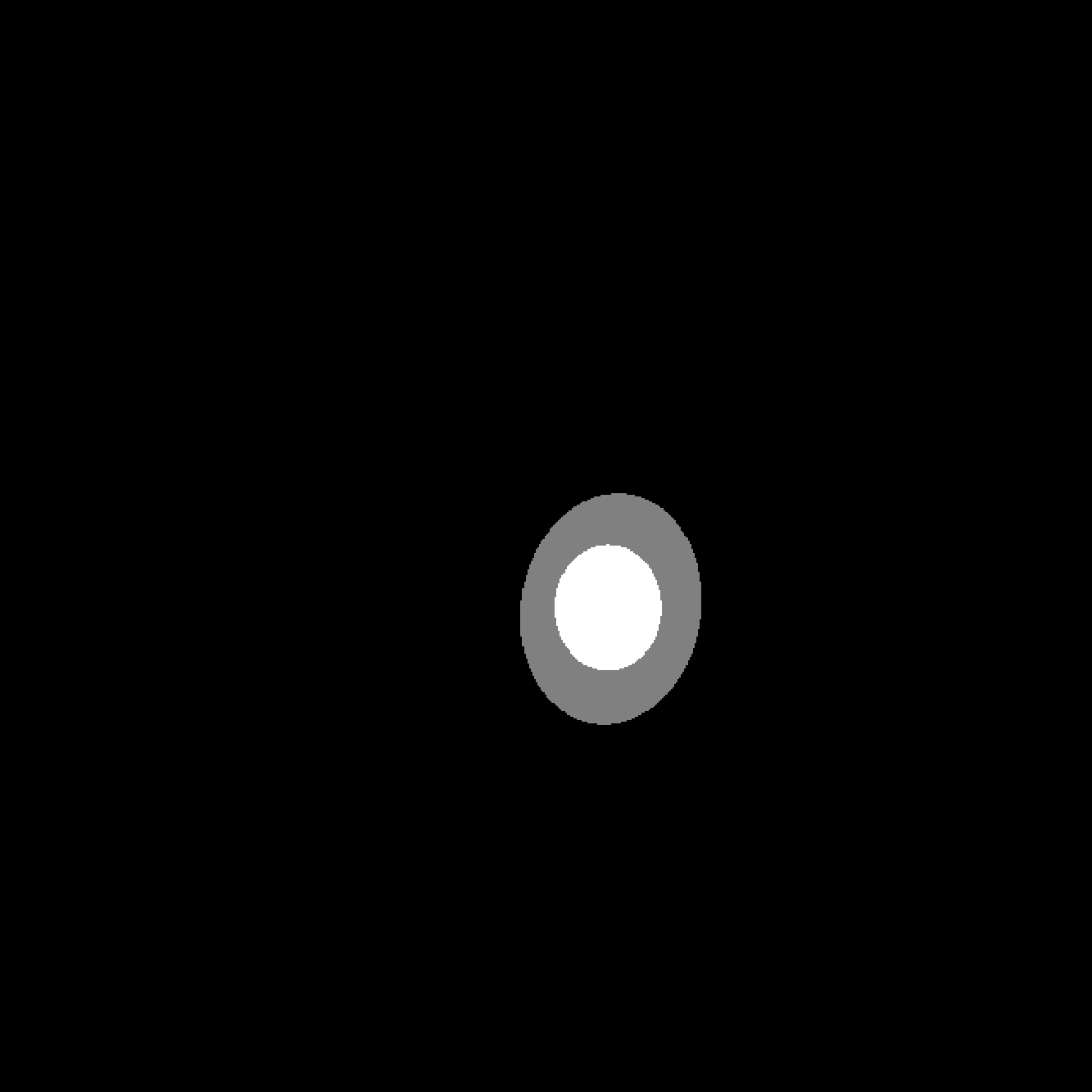}    & 
       \includegraphics[width=0.14\textwidth]{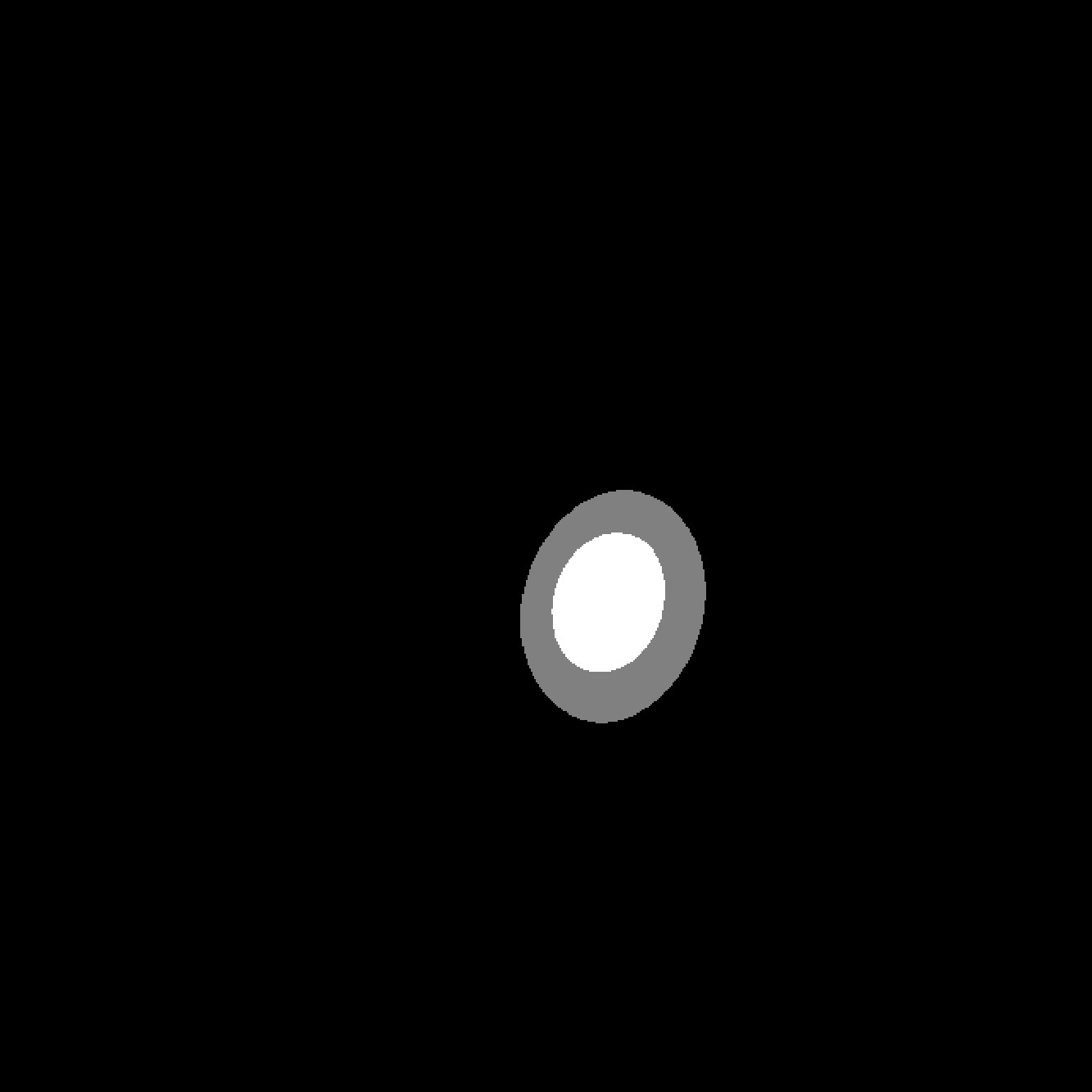}  \\       
    \end{tabular}%
    }
    \caption{Visual segmentation results on RIM-ONE dataset delivered by our network.}
  \label{DRISHTI}
\end{figure*}

\begin{figure*}
  \centering
  \resizebox{0.78\textwidth}{!}{%
  \begin{tabular}{cccc}
       \includegraphics[width=0.14\textwidth]{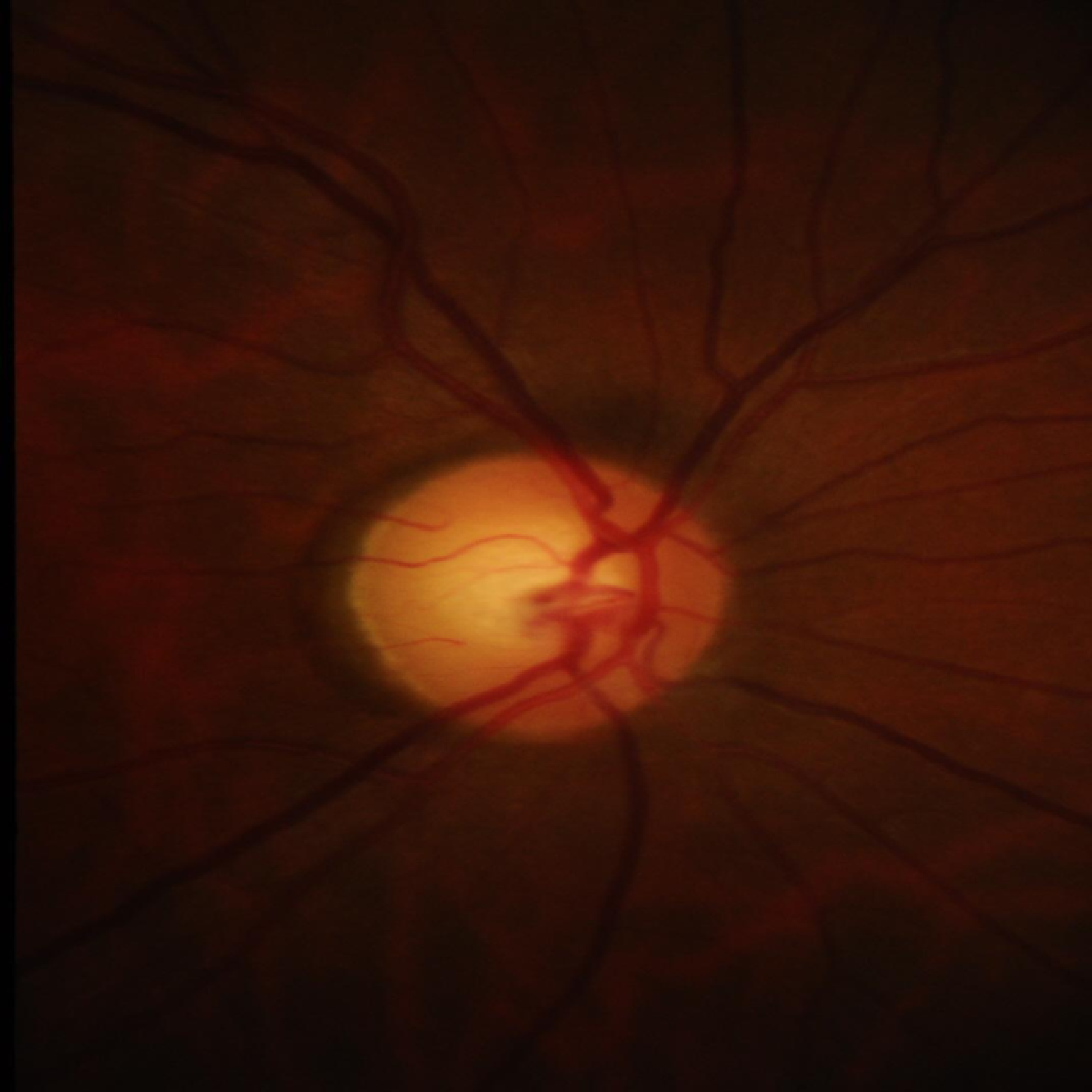}  & 
       \includegraphics[width=0.14\textwidth]{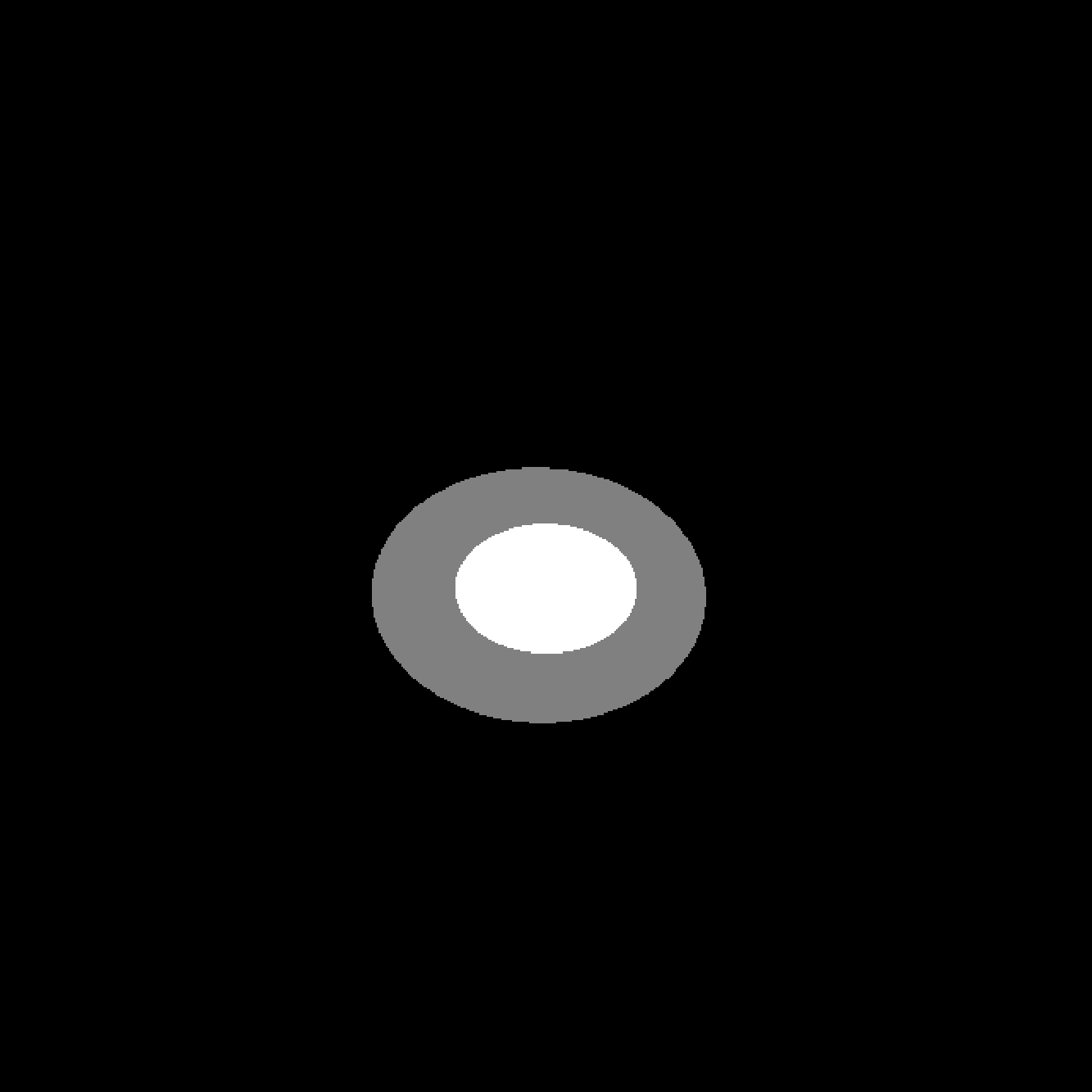}    & 
       \includegraphics[width=0.14\textwidth]{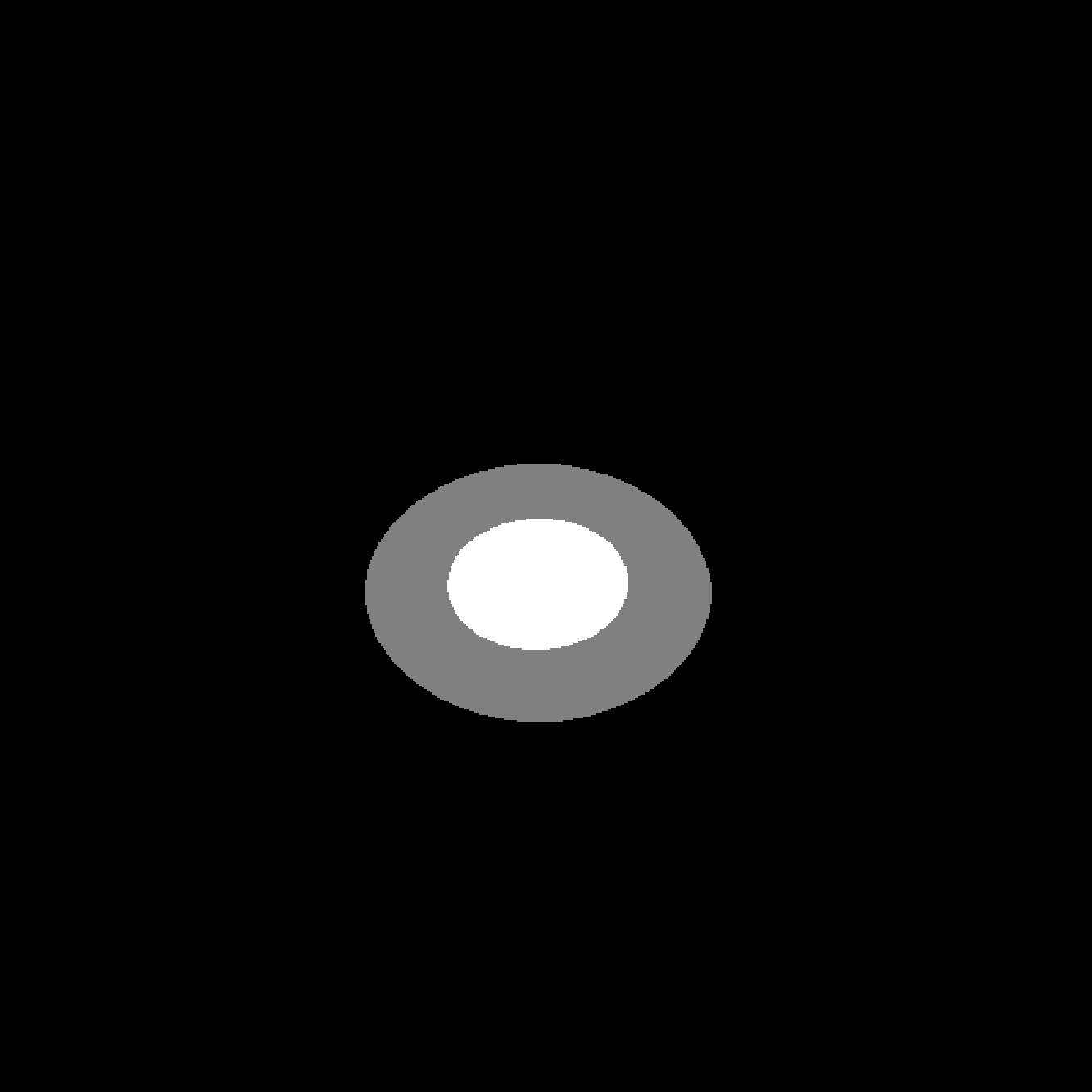}  \\
       \includegraphics[width=0.14\textwidth]{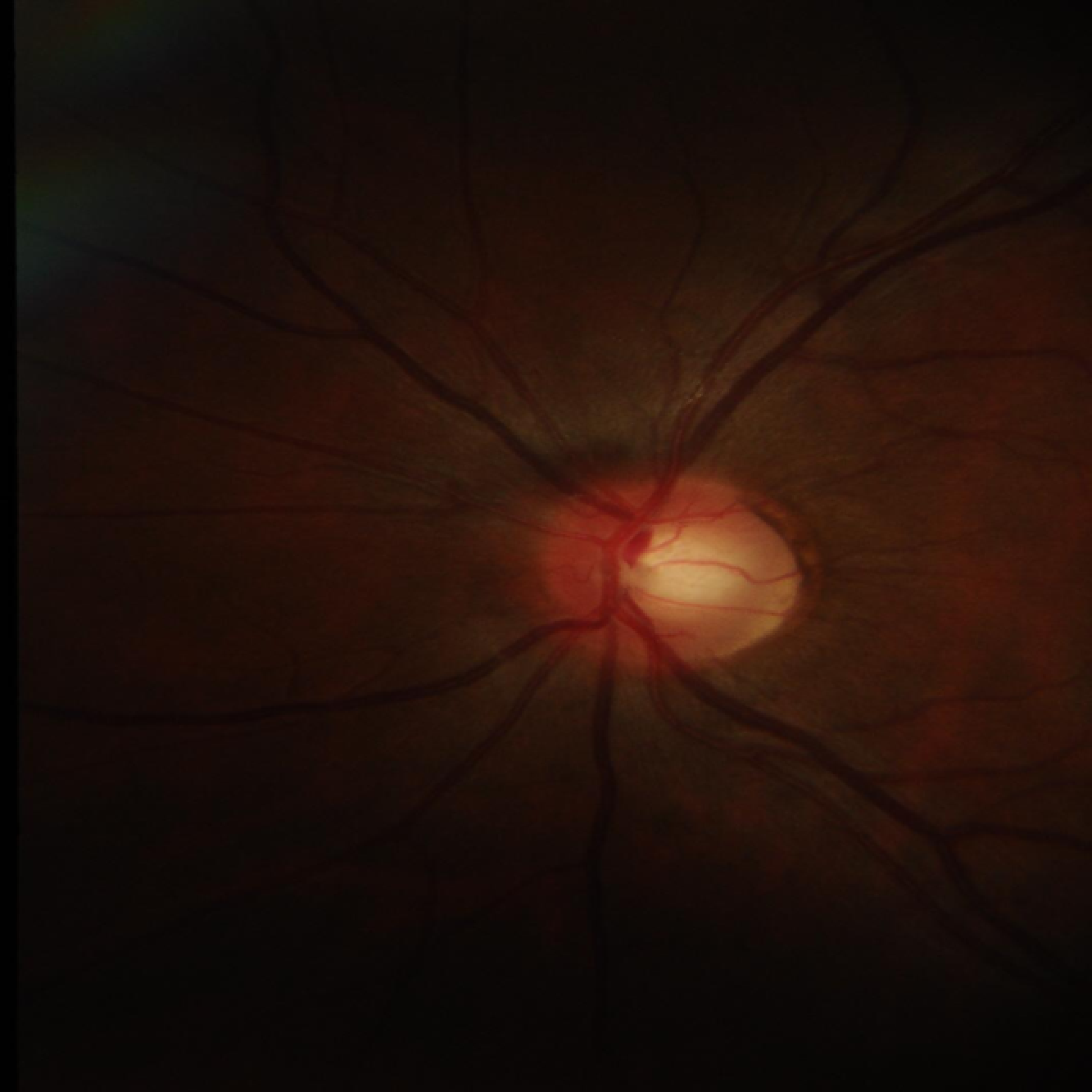}  & 
       \includegraphics[width=0.14\textwidth]{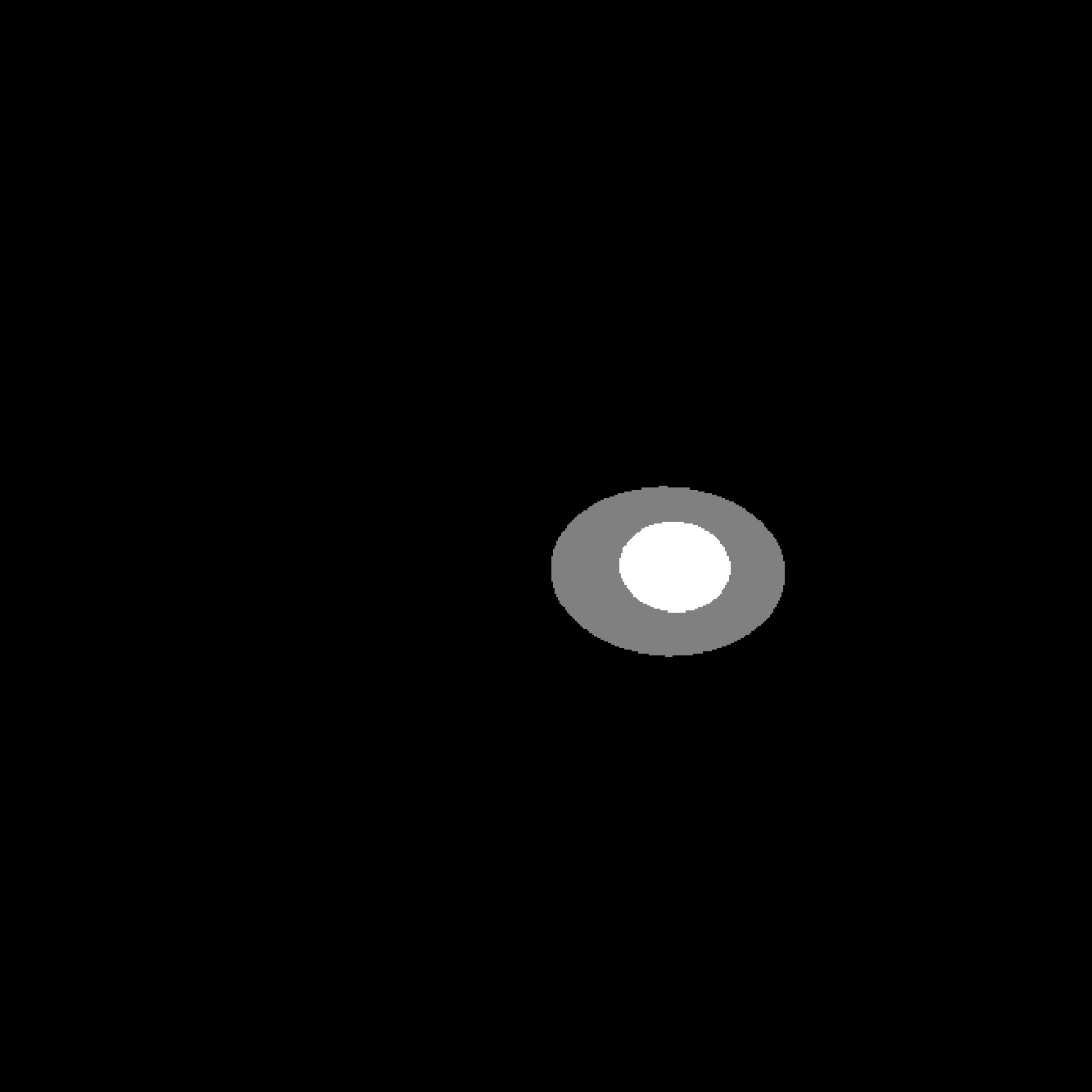}    & 
       \includegraphics[width=0.14\textwidth]{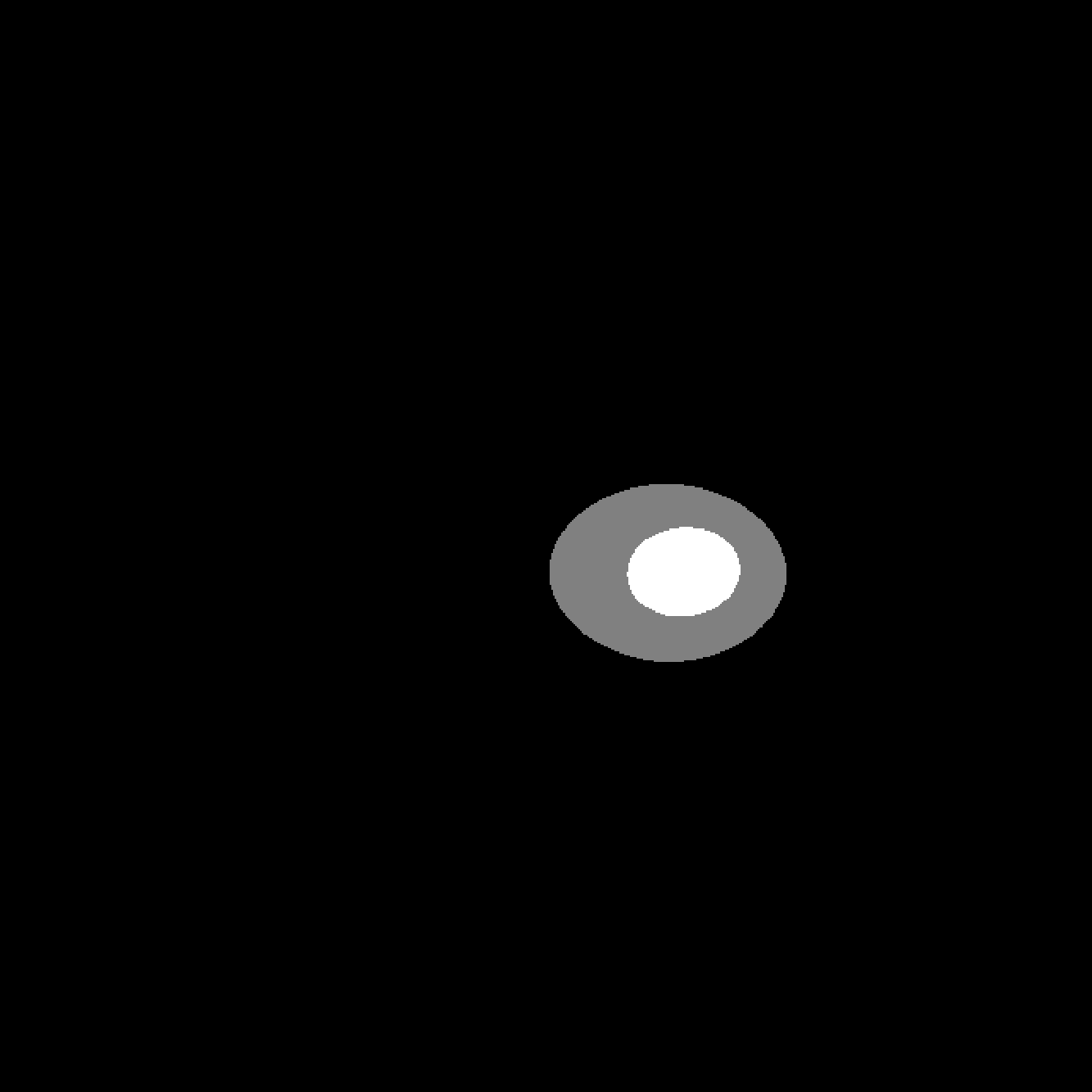}  \\       
        \includegraphics[width=0.14\textwidth]{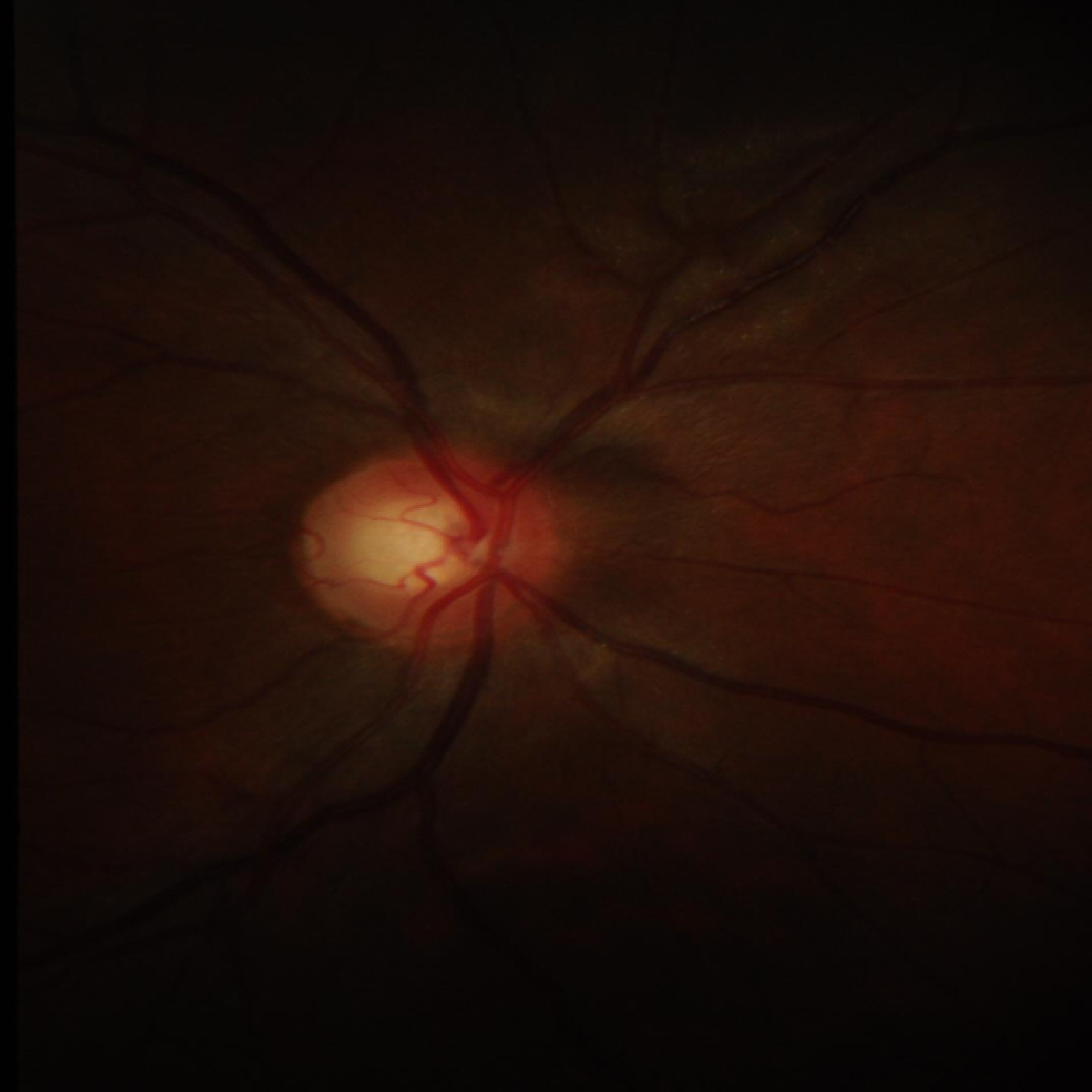}  & 
       \includegraphics[width=0.14\textwidth]{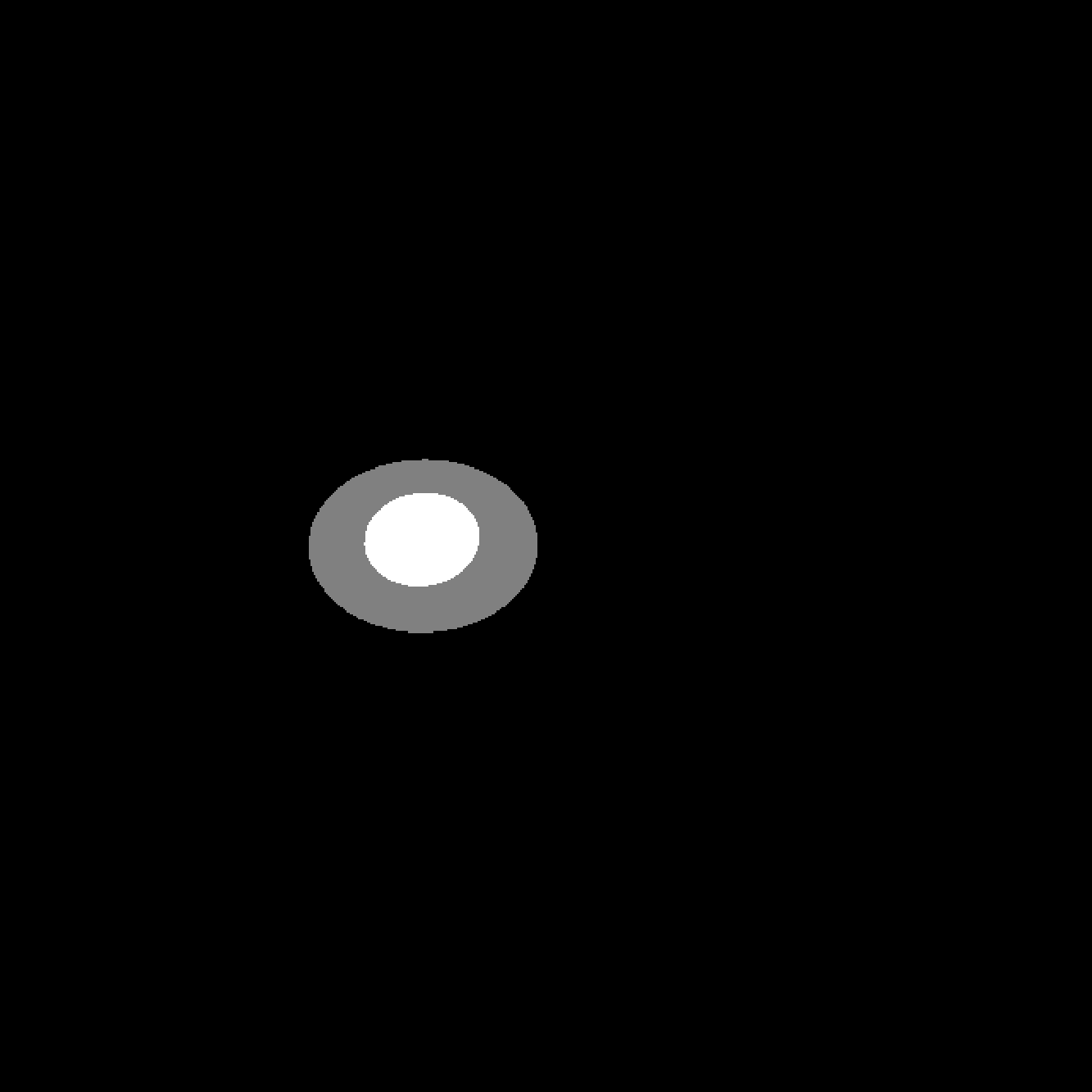}    & 
       \includegraphics[width=0.14\textwidth]{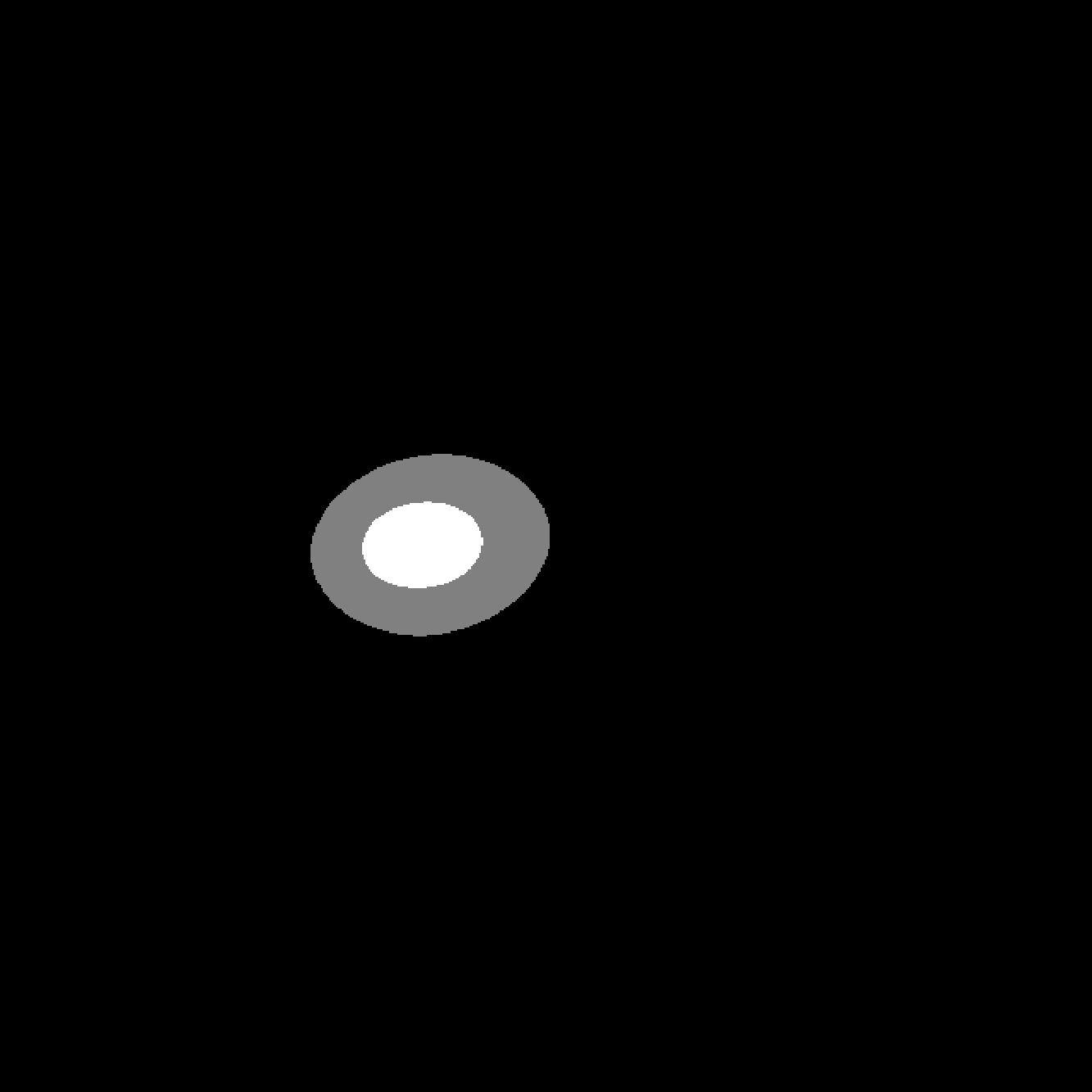}  \\  
        \includegraphics[width=0.14\textwidth]{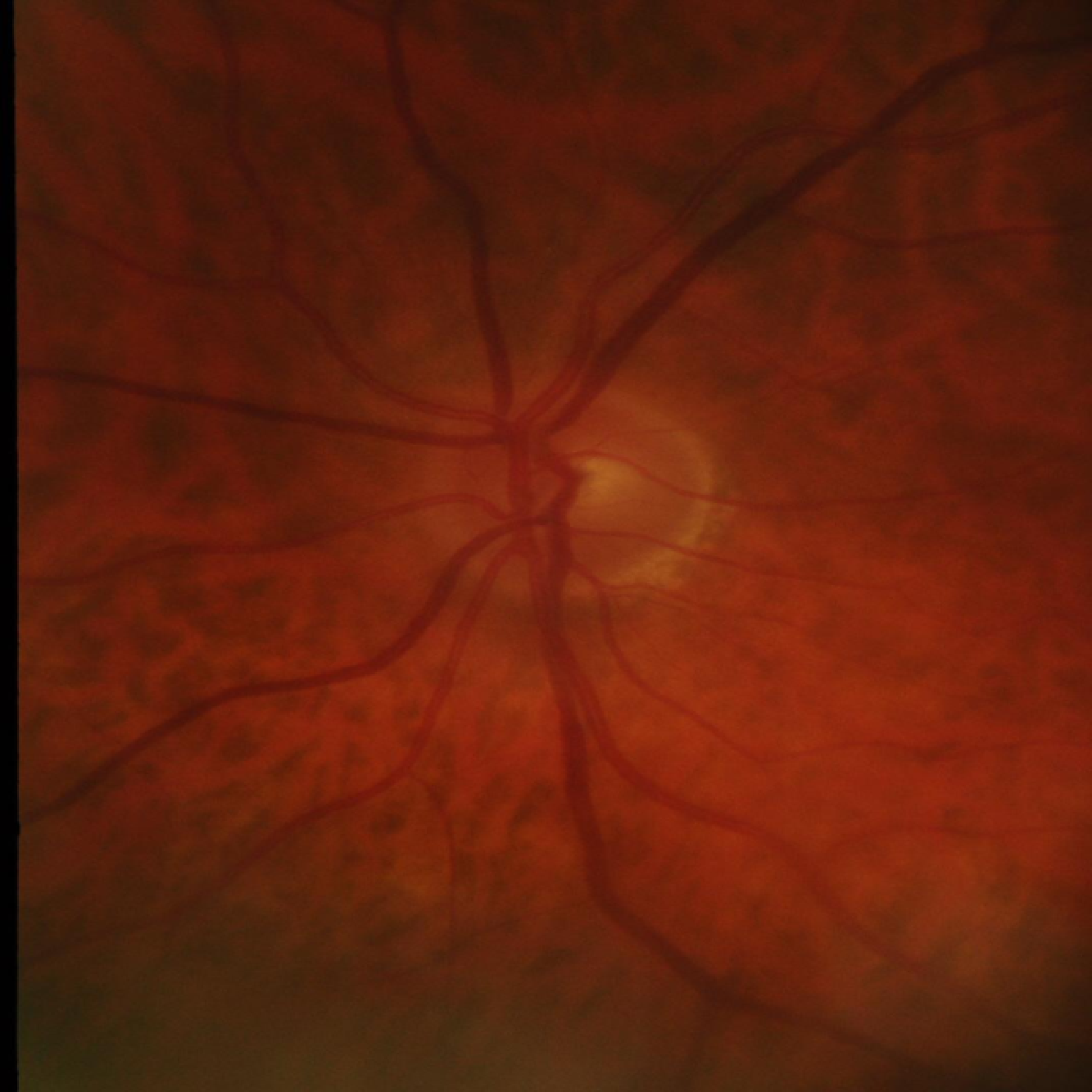}  & 
       \includegraphics[width=0.14\textwidth]{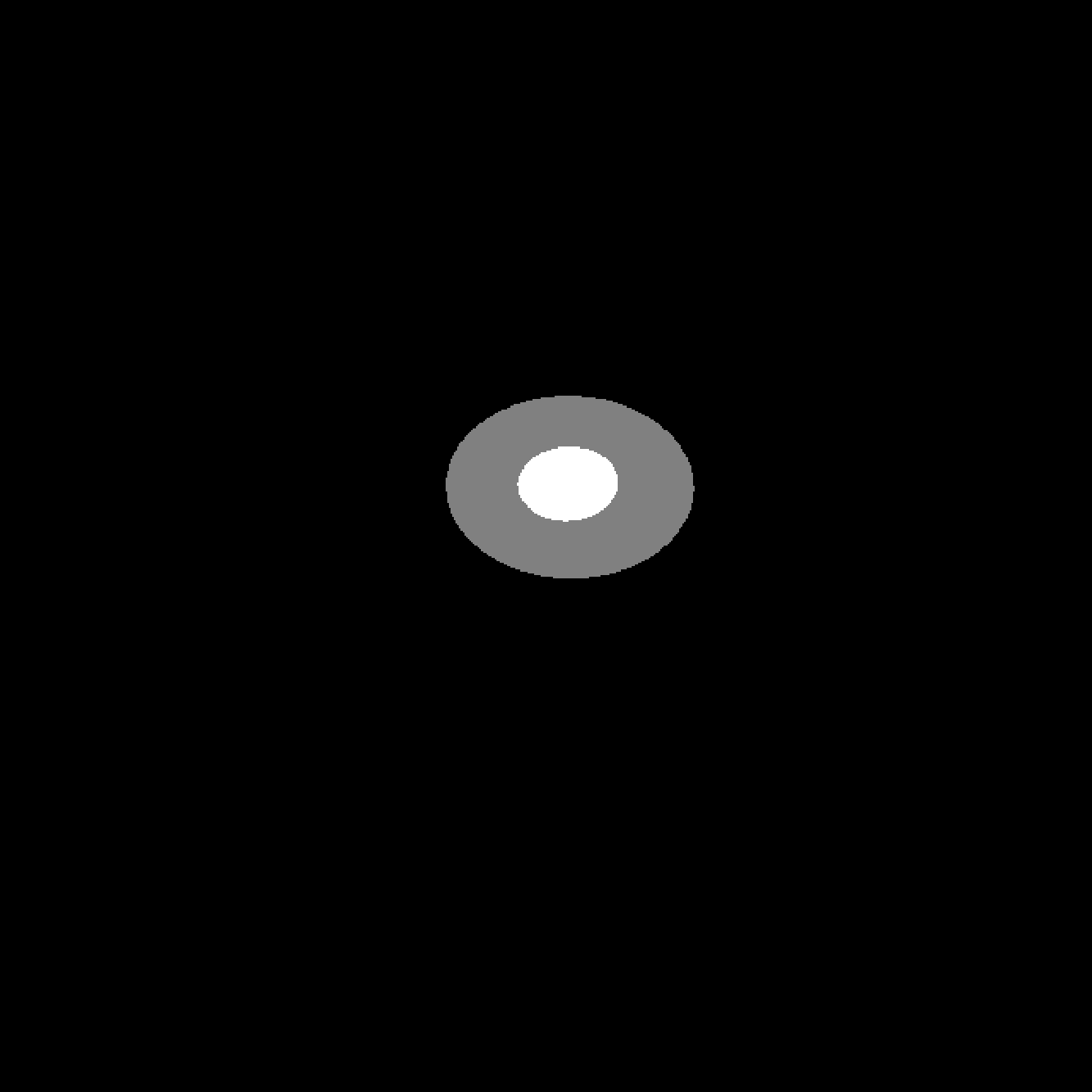}    & 
       \includegraphics[width=0.14\textwidth]{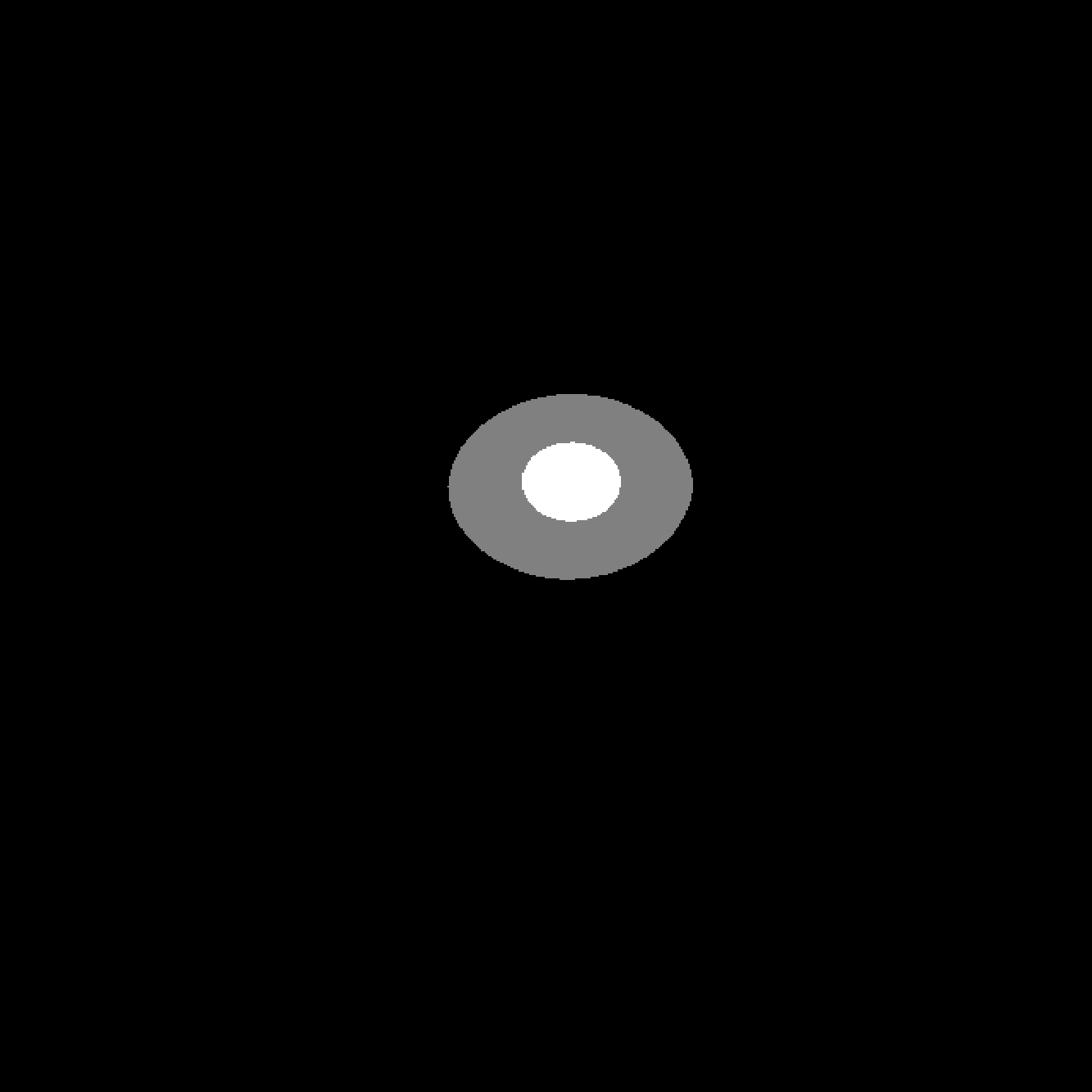}  \\  
    \includegraphics[width=0.14\textwidth]{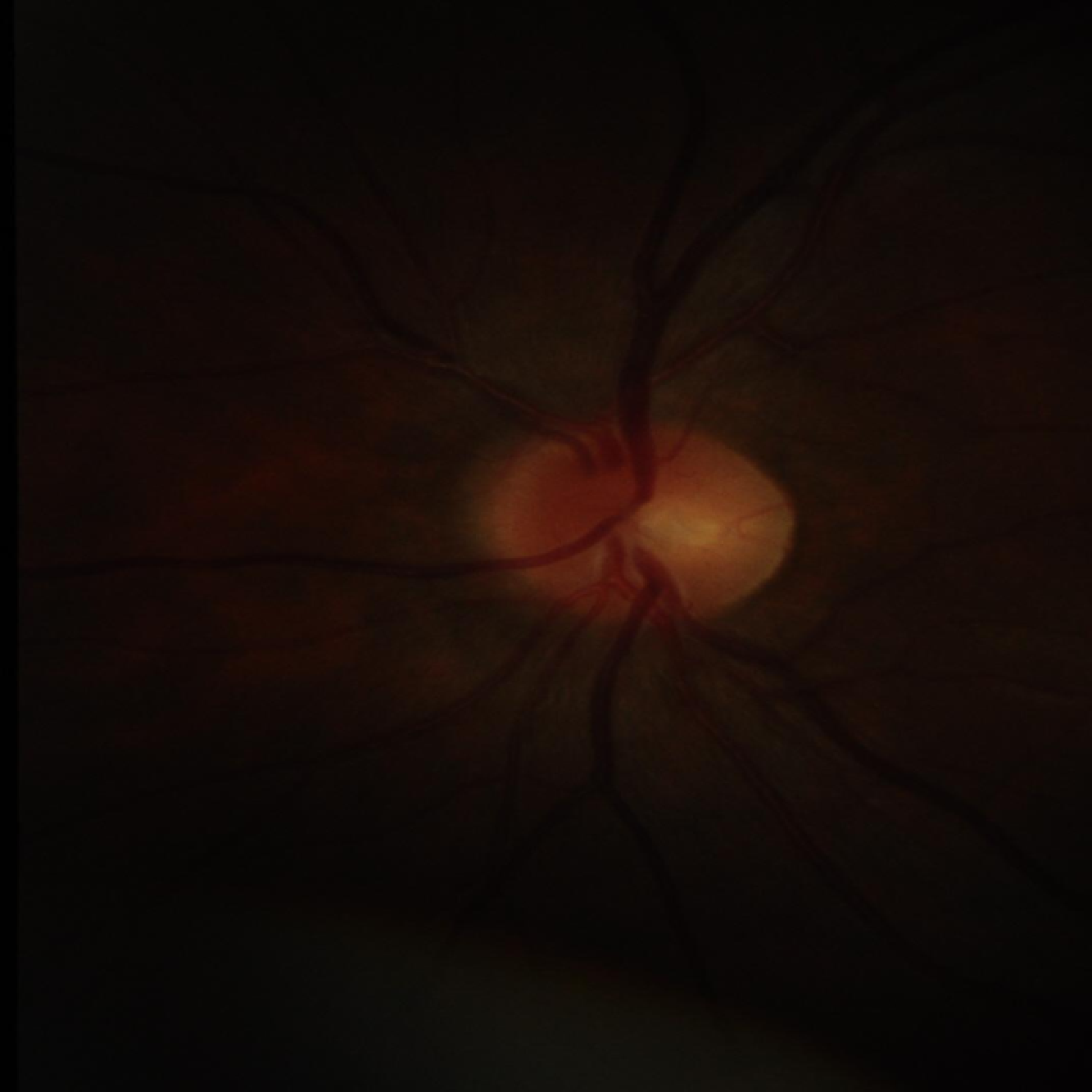}  & 
       \includegraphics[width=0.14\textwidth]{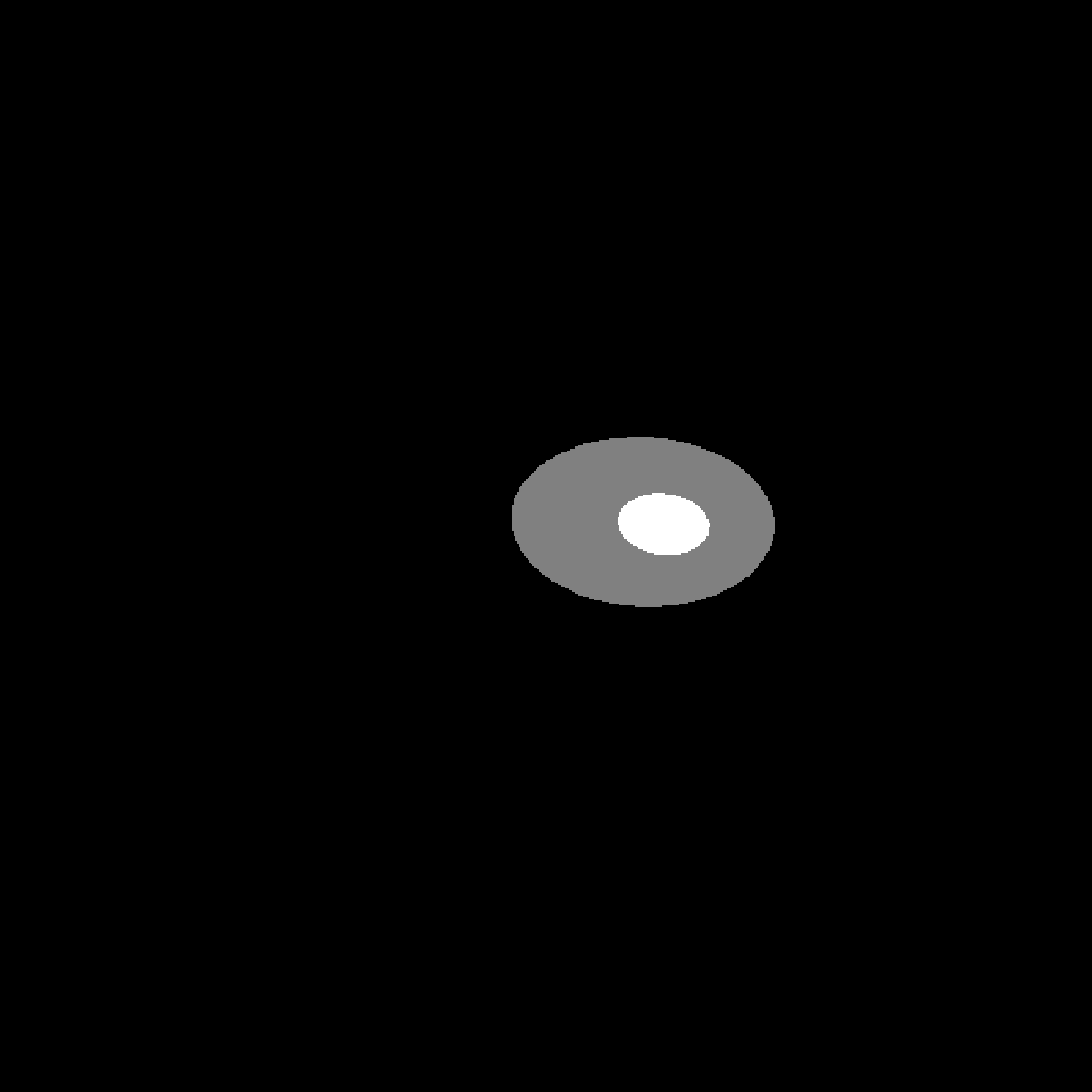}    & 
       \includegraphics[width=0.14\textwidth]{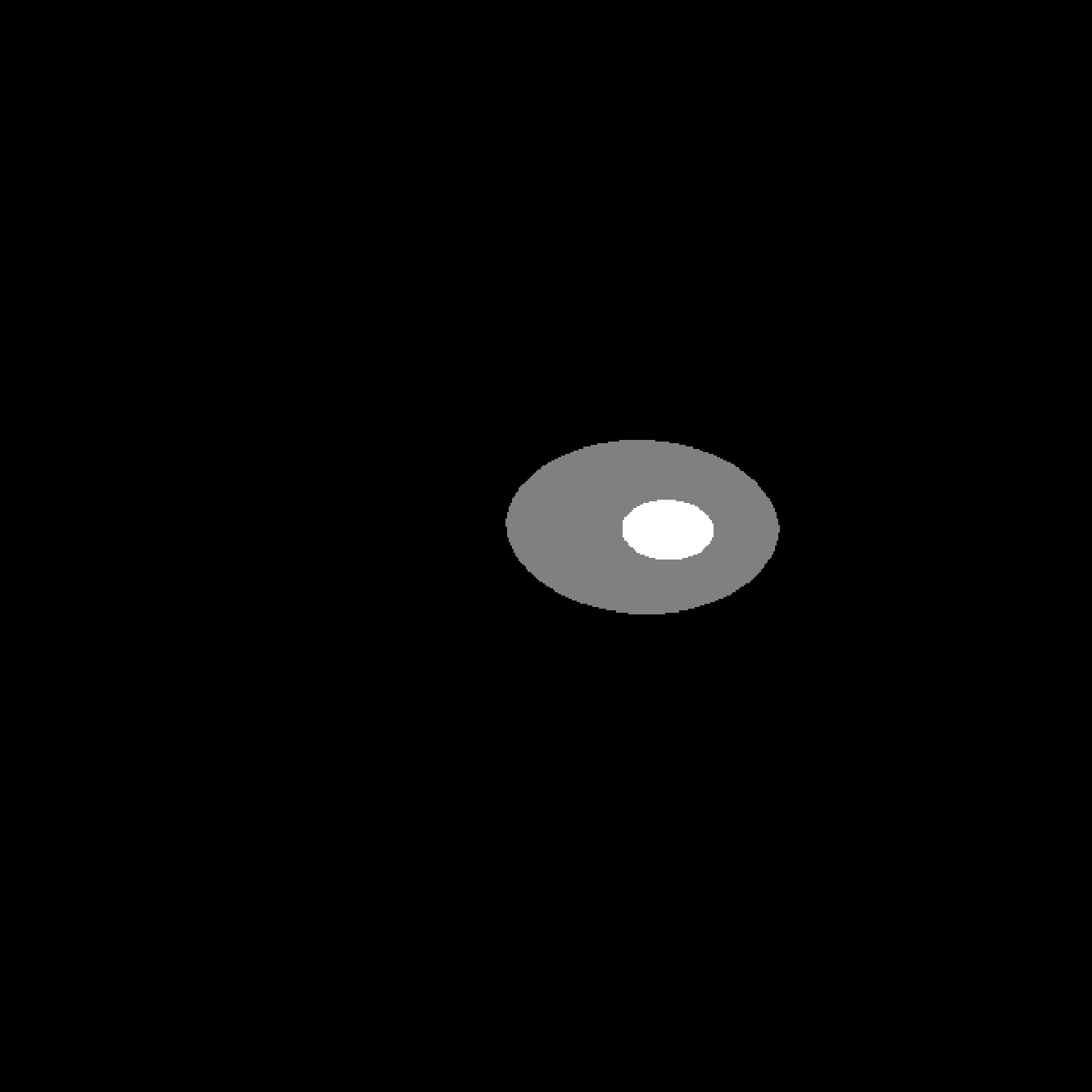}  \\  
    \end{tabular}
    }
    \caption{Visual segmentation results on DRISHTI-GS dataset delivered by our network.}
  \label{RIM-ONE}
\end{figure*}

After the compilation of results, obtained by the proposed design, comparison with state-of-the-art techniques is performed in this section. 
The visual results of our simulation on the three datasets are shown in Figures \ref{DRISHTI} and \ref{RIM-ONE} respectively. 
In each figure, moving from left to right, the first column shows the original images, the second column shows the ground truth images and the third column shows the segmented images. 

\begin{table*}[t]
\centering
\caption{Performance comparison of the developed model for OD/OC detection of Drishti dataset(\%).} 
\scalebox{1}{
		\resizebox{1\textwidth}{!}{
			\setlength\extrarowheight{3pt}
\begin{tabular}{@{\extracolsep{5pt}}llcccccccc@{}}
\toprule

\multicolumn{ 1}{l}{{\bf Author}} & \multicolumn{ 1}{l}{{\bf Method}} &  \multicolumn{ 4}{c}{{\bf OC }} & \multicolumn{ 3}{c}{{\bf OD}}\\
\cline{3-6}
\cline{7-10}

\multicolumn{ 1}{c}{{\bf }} & \multicolumn{ 1}{c}{{\bf }} & {\bf DC(F1)} & {\bf JC(O)} & {\bf Sensitivity} & {\bf Specificity} & {\bf DC(F1)} & {\bf JC(O)} & {\bf Sensitivity} & {\bf Specificity} \\
\midrule
Sevastopolsky \cite{Sevastopolsky2017} & Modified U-Net CNN  &      85.21 &      75.15 &      84.76 &      98.81 &      90.43 &       83.5 &      91.56 &      99.69 \\

Zilly et al. \cite{zilly2017glaucoma} & Ensemble CNN  &       87.1 &         85 &          - &          - &       97.3 &       91.4 &          - &          - \\

Sedai et al. \cite{Sedai16-2} & Regression based method &         85 &          - &          - &          - &         95 &          - &          - &          - \\

Zhou et al. \cite{zhou2019optic} & Locally statistical active contour model &       84.7 &          - &          - &          - &       95.5 &          - &          - &          - \\

Son et al.\cite{son2019towards} &  GAN Network &      86.43 &      77.48 &      85.39 &      99.07 &      95.27 &      91.85 &      97.47 &      99.77 \\

Chakravarty et al. \cite{ChakravartyMay} & RNN based RACE-net &         87 &          - &          - &          - &         97 &          - &          - &          - \\

Fu et al. \cite{Fu2018a} &      M-net &      86.18 &       77.3 &      88.22 &      98.62 &      96.78 &      93.86 &      97.11 &      99.91 \\

Wang et al. \cite{Wang2019} & Patch based Output Space Adversarial Learning &       85.8 &          - &          - &          - &       96.5 &          - &          - &          - \\

Gu et al. \cite{GuOct.} & Encoder-Decoder CE-Net &      88.18 &      80.06 &      88.19 &      99.09 &      96.42 &      93.23 &      97.59 &       99.9 \\

Xu et al. \cite{xu2019mixed} & U-shaped convolutional neural network &       89.2 &      82.30 &          - &          - &       97.8 &       94.9 &          - &          - \\

{\bf \textsc{Proposed}} & {\bf \textsc{EDDense-Net}} &       92.26 &      86.25 &      96.04 &      99.78 &      96.10 &      92.90 &      95.85 &      99.87 \\
\bottomrule
\end{tabular}
}}
\label{OD_OC_Drishti}
\end{table*}

\begin{table*}[httbp]
\centering
\caption{Performance comparison of the developed model for the tasks of OD and OC segmentation based on RIM-ONE dataset(\%).}
\scalebox{1}{
		\resizebox{1\textwidth}{!}{
			\setlength\extrarowheight{3pt}
\begin{tabular}{@{\extracolsep{5pt}}llcccccccc@{}}
\toprule

\multicolumn{ 1}{c}{{\bf Author}} & \multicolumn{ 1}{c}{{\bf Method}} &   \multicolumn{ 4}{c}{{\bf OC }} &             \multicolumn{ 4}{c}{{\bf OD}} \\
\cline{3-6}
\cline{7-10}

\multicolumn{ 1}{c}{{\bf }} & \multicolumn{ 1}{c}{{\bf }} & {\bf DC(F1)} & {\bf JC(O)} & {\bf Sensitivity} & {\bf Specificity} & {\bf DC(F1)} & {\bf JC(O)} & {\bf Sensitivity} & {\bf Specificity} \\
\midrule
Cheng et al. \cite{ChengJune} & Superpixel classification &       74.4 &       73.2 &          - &          - &       89.2 &      82.93 &          - &          - \\

Aquino et al. \cite{aquino2010detecting} & Template-based approach &          - &          - &          - &          - &       90.1 &       84.2 &          - &          - \\

Arnay et al. \cite{Arnay2017} & Ant colony optimization &          - &       75.7 &          - &          - &          - &          - &          - &          - \\

Sevastopolsky \cite{Sevastopolsky2017} & Modified U-Net CNN &         82 &         69 &      75.45 &      99.76 &         95 &         89 &      95.02 &      99.73 \\

Zilly et al. \cite{zilly2017glaucoma} & Ensemble CNN  &       82.4 &       80.2 &          - &          - &       94.2 &         89 &          - &          - \\

Zhou et al. \cite{zhou2019optic} & Locally statistical Active contour model &       78.5 &          - &          - &          - &       85.3 &          - &          - &          - \\

Son et al.\cite{son2019towards} &  GAN Network &       82.5 &      71.65 &      81.42 &      99.65 &      95.32 &      91.22 &      94.57 &      99.87 \\

Fu et al. \cite{Fu2018a} &      M-net &      83.48 &         73 &      81.46 &      99.67 &      95.26 &      91.14 &      94.81 &      99.86 \\

Gu et al. \cite{GuOct.} & Encoder-Decoder CE-Net &      84.35 &      74.24 &      83.52 &       99.7 &      95.27 &      91.15 &      95.02 &      99.86 \\

Wang et al. \cite{Wang2019} & Patch based Output Space Adversarial Learning &       78.7 &          - &          - &          - &       86.5 &          - &          - &          - \\

Xu et al. \cite{xu2019mixed} & U-shaped convolutional neural network &      85.64 &      75.86 &      85.15 &      99.71 &      95.61 &      91.72 &      95.21 &      99.87 \\

{\bf \textsc{Proposed}} & {\bf \textsc{EDDense-Net}} &     90.64 &      83.18 &      93.42 &      99.89 &      95.84 &      92.33 &      95.78 &      99.84 \\
\bottomrule
\end{tabular}
}}
\label{OD_OC_RIMONE}
\end{table*}

The segmentation performance of the developed model for the tasks of OD and OC detection using Drishti-GS and RIM-ONE for the evaluation metrics of error and balance accuracy (Eq.~\ref{Error} and Eq.~\ref{BAcc}) is shown in Table~\ref{E_A_Drishti}. 
It is evident from this table that the developed model achieved better BA for both OC and OD detection based on test images from both the considered datasets. 
Furthermore, it also obtained comparable results for the error metric for both OD and OC detection on the test images from both the datasets. 

\begin{table*}[httbp]
\centering
\footnotesize
\caption{Segmentation performance of the proposed EDDense-Net on DRISHTI and RIM-ONE dataset using overlapping error (E) and balanced accuracy (BA).}
\scalebox{1}{
		\resizebox{1\textwidth}{!}{
			\setlength\extrarowheight{3pt}
\begin{tabular}{lrrrrrrrr}
\toprule
           &                \multicolumn{ 4}{c}{{\bf Drishti}} &                \multicolumn{ 4}{c}{{\bf RIM-ONE}} \\

{\bf Methods} & \multicolumn{ 2}{c}{{\bf OD}} & \multicolumn{ 2}{c}{{\bf OC }} & \multicolumn{ 2}{c}{{\bf OD}} & \multicolumn{ 2}{c}{{\bf OC }} \\
\cline{2-5}
\cline{6-9}
{} &    {\bf E} &    {\bf A} &    {\bf E} &    {\bf A} &    {\bf E} &    {\bf A} &    {\bf E} &    {\bf A} \\
\midrule
\multicolumn{1}{l}{{\bf Variants proposed in ~\cite{shankaranarayana2019fully}}} &       &       &       &       &       &     &      &     \\

Model using residual blocks without depth (ResUnet) &      0.089 &      0.968 &      0.283 &      0.926 &       0.06 &      0.974 &      0.321 &      0.914 \\

Model using DRI blocks without depth (DriUnet) &      0.074 &      0.972 &      0.286 &      0.932 &       0.06 &      0.974 &      0.284 &      0.925 \\

Model using CRF post-processing and DRI blocks (DriUnet CRF) &      0.077 &      0.971 &       0.27 &      0.941 &      0.061 &      0.974 &      0.285 &      0.924 \\

Residual blocks and depth based model (Dept ResUnet) &      0.073 &      0.961 &      0.268 &      0.937 &      0.059 &      0.974 &      0.299 &     0.9353 \\

DRI blocks and Depth based model (Depth DriUnet) &      0.068 &      0.964 &      0.276 &      0.936 &      0.059 &      0.975 &       0.31 &       0.91 \\

DRI blocks and Pseudo-depth based model (PD DriUnet) &      0.071 &      0.972 &       0.24 &      0.941 &      0.058 &      0.975 &      0.284 &       0.92 \\

\multicolumn{1}{l}{{\bf Proposed}} &       &       &       &       &       &     &      &     \\

{\bf EDDense-Net} &     0.0709 &     0.9585 &     0.1375 &     0.9604 &     0.0691 &     0.9578 &     0.1690 &     0.9342 \\
\bottomrule
\end{tabular}
}}
\label{E_A_Drishti}
\vspace{1cm}
\end{table*}

For visual analysis, the OD and OC segmentation results of the developed model for RIM-ONE and DRISHTI-GS datasets are shown in Figure~\ref{RIM-ONE} and Figure~\ref{DRISHTI} respectively. 

To evaluate and compare our results with those of state-of-the-art models, we have presented and summarized the results in tabular forms.  
As given in the tables, the Se, Sp, and Accuracy of our models are 0.8144, 0.9843, and 0.9726 respectively.

In Table~\ref{OD_OC_Drishti}, the results of our proposed model, implemented on the Drishti dataset, are compared with those of state-of-the-art. 
Se, Sp, and Accuracy of our model are 0.8286, 0.9824, and 0.9689 respectively.

Similarly, results achieved from the implementation of our model on the RIM-ONE dataset are compared in Table~\ref{OD_OC_RIMONE}. 
From this experiment, dice, Jaccard Se, and Sp of our model are 0.9226, 0.8625,0.9604, and 0.9978 respectively for OC segmentation and 0.9610, 0.9290, 0.9585, and 0.9987 respectively for OD segmentation.

From the comparisons with state-of-the-art, it is obvious that our proposed model outperformed other existing models with respect to well-known evaluation metric accuracy on three well-known and publicly available datasets.

Figure~\ref{DRISHTI} and Figure~\ref{RIM-ONE} represent OD and OC segmentation results on the DRISHTI-GS and RIM-ONE datasets respectively.
In~\cite{Sevastopolsky2017}, Sevastopolsky et al. proposed a method based on the U-Net convolutional network to segment out the OD and OC. 
In the case of OD, the achieved dice and Jaccard are 90.43\% and 89\%. Whereas for OC, it is 82\% and 69\% on Rim-one. 
For OC, attained dice and Jaccard are 85\% and 75\%, and for OD values are 95\% and 89\% respectively on the Drishti dataset.
Zilly et al. proposed an ensemble learning-based CNN model and to reduce the computational complexity, entropy sampling is performed and a framework is designed using boosting for convolutional layers. 
This algorithm achieved dice and Jaccard of 94.2\% and 89\% for OD 82.41\% and 80.2\% for OC on Rim-one. 
In the case of the Drishti dataset, the acquired dice and Jaccard for OD are 97.3\% and 91.4\% and for
OC; dice and Jaccard are 87.1\% and 85\% respectively \cite{zilly2017glaucoma}.
Cheng et al. presented a superpixel based classification model~\cite{ChengJune} for the segmentation purpose and the achieved dice and Jaccard are 89.2\% and 82.93\%, and in the case of OC, 74.4\% and 73.2\% respectively which is significantly lower than our proposed method. 
Aquino et al. performed the segmentation of OD using a template-based method \cite{aquino2010detecting}. 
In this method, morphological and edge detection-based techniques were used; followed by CHT which is used to detect the boundary of OD. 
The attained dice and Jaccard are 90.1\% and 84.2\% respectively. 
Ant Colony optimization method was used by~\cite{Arnay2017} for OC segmentation. 
This method achieved a lower Jaccard value of 75.7\% for cup segmentation. 
Sedai et al. proposed a regression-based method for the segmentation of the OD and OC \cite{Sedai16-2}. 
Initially, CHT is employed to get the OD and then it is used to estimate the shapes of OD and OC. 
For OD, the acquired dice is 95\% and for OC the dice is 85\%. Zhou et al. presented a locally statistically active contour model with the structure prior (LSACM-SP) approach for joint segmentation~\cite{zhou2019optic}. 
They achieved 95.5\% and 84.7\% dice values for disc and cup on Drishti and 85.3\% and 78.5\% values on rim-one. 
RNN architecture (RACE-net) has achieved 97\% and 87\% values for dice on disc and cup for Drishti~\cite{ChakravartyMay}. 
Wang et al. have proposed a patch-based adversarial network and have achieved 85.8\% and 96.5\% dice for OC and OD respectively for Drishti and 78.7\% and 86.5\% cup and disc dice values on rim-one dataset~\cite{Wang2019}. 
Their dice value for the cup is much lower than achieved by the proposed architecture. 
Xu et al. have designed a U-shaped convolutional neural network with multi-scale input and multi-kernel modules (MSMKU) for disc and cup segmentation but their results were lower than the proposed model.  
Our method has achieved state-of-the values for Jaccard on disc and cup when compared with other methods.
Table \ref{OD_OC_Drishti} and \ref{OD_OC_RIMONE} 4 also give a comparison of these techniques with the proposed model on both datasets.
We have also evaluated our model in terms of sensitivity and specificity. 
The proposed model achieves state-of-the-art performance on both datasets for disc and cup. 
It achieves sensitivity and specificity of 93.42\%, 99.89\% respectively for the OC, and 95.78\% and 99.84\% for the OD segmentation on RIM-ONE. 
On Drishti, it achieves sensitivity and specificity of 96.04\% and 99.78\% for cup and 95.85\% and 99.87\% for OD segmentation which are higher than other state-of-the-art methods.

For the OD and OC, the performance of the proposed method was compared with a recent work~\cite{Shankaranarayana2017} in terms of overlapping error (E) and balance accuracy (A) in Table 5 on the DRISHTI-GS and RIM-ONE datasets. 
In~\cite{Shankaranarayana2017}, the authors proposed a fully convolutional network (FCN) for the depth estimation using a pre-training scheme called pseudo-depth reconstruction and a dilated residual inception (DRI) block for multi-scale extraction of features. 
The guided network was used for OD and OC segmentation using a depth map as a guide and evaluated E and A values for different experiments as shown in Table \ref{E_A_Drishti}.
It can be observed from the results that the proposed EDDense-Net outperforms all the variants reported in~\cite{Shankaranarayana2017} in terms of E and A measures for the OC segmentation task. 
Also, the proposed method performs better than all the compared methods on the task of OD segmentation in terms of A, with comparable performance in terms of E.  
As shown in Table \ref{E_A_Drishti}, our model has achieved state-of-the-art performance with E and A as 0.0709 and 0.95850 for OD and 0.1375 and 0.9604 for OC respectively on Drishti and 0.0691 and 0.9578 for OD as well as 0.1690 and 0.9342 for OC on RIM-ONE.

\section{Conclusion}
Glaucoma, after cataracts, is the world's second leading cause of permanent blindness.
In order to avoid permanent blindness, it is critical to diagnose glaucoma early.
We presented an EDDense-Net segmentation network for the joint segmentation of OC and OD in order to proceed with the diagnosis of this disease.
The network was able to gather and transmit spatial information from the image by concatenating features in the dense layers.
The edge information transfer from encoder layers to decoder layers is used to achieve network convergence quickly.
EDDense-Net has four primary design characteristics: EDDense-Net, for starters, uses thick routes for feature empowerment, which aids in the extraction of minor information from the image.
Second, to reduce spatial information loss, the optimal number of filters in all convolution layers was utilized.
Third, utilizing grouped convolution reduces the complexity even more.
Finally, instead of using simple pixel classification, which uses generalized Dice loss to tackle the class imbalance problem in semantic segmentation, dice pixel classification is employed in the last layer. 
On two publicly available datasets, the proposed network outperformed existing state-of-the-art approaches in terms of accuracy and efficiency.
For the diagnosis and analysis of glaucoma, this method can be utilized as a second opinion system to assist medical ophthalmologists.
We will enhance the accuracy of OC and OD segmentation in the future, as well as take into account other retinal illnesses.

 \bibliographystyle{elsarticle-num} 
\bibliography{Manuscript}





\end{document}